\documentclass[aps,prb,twocolumn,showpacs,superscriptaddress,groupedaddress]{revtex4}  % for review and submission
\usepackage{graphicx}  % needed for figures
\usepackage{dcolumn}   % needed for some tables
\usepackage{bm}        % for math
\usepackage{amssymb}   % for math
\usepackage{amsmath}
\usepackage[titletoc,title]{appendix}
\usepackage{array}
\usepackage{tabularx}
\begin{document}

\title{External Modulation and Switching of Acoustic Phonons: Comparative Roles of Potential Distributions}

\author{H. \surname{Jeong}}

\affiliation{School of Information and Communications, Gwangju Institute of Science and Technology,
Gwangju 500-712, Korea}

\author{Y. D.  \surname{Jho}}
\email{jho@gist.ac.kr} \affiliation{School of Information and Communications, Gwangju Institute of
Science and Technology, Gwangju 500-712, Korea}

\author{S. H. \surname{Rhim}}
\affiliation{Department of Physics and Energy Harvest Storage Research Center, University of Ulsan, Ulsan 680-749, Korea}

\author{K. J.  \surname{Yee}}
\affiliation{Department of Physics, Chungnam National University, DaeJeon 305-764, Korea}

\author{J. P. \surname{Shim}}

\affiliation{School of Information and Communications, Gwangju Institute of Science and Technology,
Gwangju 500-712, Korea}

\author{D. S. \surname{Lee}}

\affiliation{School of Information and Communications, Gwangju Institute of Science and Technology,
Gwangju 500-712, Korea}

%\author{P. \surname{Gr\"{u}nberg}}
%\affiliation{Department of Nanobio Materials and Electronics, Gwangju Institute of Science and Technology,
%Gwangju 500-712, Korea}

\author{J. W. \surname{Ju}}
\author{J. H. \surname{Baek}}
\affiliation{LED Process Technology Center, KOPTI, Gwangju 500-779, Korea}

\author{C. J. \surname{Stanton}}
\affiliation{Department of Physics, University of Florida, Gainesville, Florida 32611, USA}

\date{\today}

\begin{abstract}

Acoustic phonons can be coherently generated by ultrafast displacive screening of potential gradients, often enhanced by the strong built-in piezoelectric fields, in wurtzite semiconductors. In such structures, transverse symmetry within the \textit{c} plane hinders both the generation and detection of the transverse acoustic (TA) modes, and only longitudinal acoustic (LA) mode is generated. We show that even for $c$-GaN, the application of asymmetric potential distributions in the \textit{c} plane can break the symmetry and selection rules, thus switching on the normally forbidden TA mode. This is in contrast to the LA mode, the strength of which varies with the symmetric potential distributions. By comparing transient differential reflectivity spectra in structures with and without asymmetric potential distributions, the role of the electrically attained anisotropy was further revealed by the digitized appearance of the TA mode, in clear contrast to the monotonic LA mode, and by modulations in the  propagation velocities, optical birefringence, and geometrically varying sensitivities, the underlying mechanisms of which are modeled by electric-field-dependent perturbations of the dielectric tensors, incorporating the results of elastic modulations.
\end{abstract}

%This is in contrast to the LA mode, the strength of which varies with the vertical but not the lateral bias.

%\keywords{GaN, acoustic phonon, nanophononic waves, THz}

 \maketitle

%(previously used pacs numbers in PRL)
%42.50.Wk Mechanical effects of light on material media, microstructures and particles
%63.20.-e Phonons in crystal lattices
%78.20.hc Laser ultrasonics
%78.35.+c Brillouin and Rayleigh scattering; other light scattering

\section{Introduction}

Restructuring materials below the acoustic (AC) phonon mean free path or implementing variations in their elastic properties suggested new perspectives regarding the controllability of the acoustic and thermal properties of crystals, such as thermal conductivity management \cite{Yu,Pernot} and charge transfer via AC pulses \cite{Young}, as well as the concept of phonon lasers\cite{Fainstein,Mahboob}. Particularly in wurtzite crystals, where pseudomorphic strain caused by lattice mismatch at the interfaces leads to large internal polarization fields, photogenerated carriers screen out the strain-induced piezoelectric field, launching coherent AC motions\cite{Chern,Sun99,JeongZno}. This effect turned out to be very effective compared to thermal strain generation\cite{Thomsen} and deformation potential coupling\cite{Wright}. By virtue of strong electromechanical coupling, AC phonons in GaN-based piezoelectric heterostructures have been further investigated in terms of the strain dependence\cite{Yahng}, tunability of zone-folded AC phonons\cite{Sun}, spatial modulation of AC waves \cite{Lin}, and terahertz (THz) electromagnetic radiation at acoustically mismatched interfaces \cite{Armstrong,Stanton2003}.

In this regard, GaN-based piezoelectric heterostructures could serve as a versatile test bed for characterizing the AC functionalities with extraordinary generation efficiency in the scheme of time-resolved ultrasonics. More specifically, (1) both zone-folded \cite{Sun,Sun99,Lin} and propagating AC phonons\cite{Yahng,Kim,Chen,Wen} could be detected; (2) because of the small absorption depth ($\leq$100 nm) compared to the conventional excitation spot size ($\geq$10 $\mu$m), the mode of the generated phonons depends on the growth direction; e.g., transverse acoustic (TA) and longitudinal acoustic (LA) phonons can appear simultaneously in the anisotropic plane \cite{Wen,Chen}, or LA phonons alone can be generated in the isotropic $c$ plane\cite{Yahng, Kim, Lin}; (3) both the deformation potential and displacive screening of electric fields, combined with the piezoelectricity, contribute to generating AC phonons having compositional amplitudes that depend on the crystal axis\cite{Chern}.

Owing to the well-established crystallinity of wurtzite semiconductors along the symmetry axis, where no shear mode can be excited by photoexcitation, and their poor detection sensitivity for TA phonons\cite{Hurley}, previous studies were rather restricted to the LA properties among the three AC polarizations, although the TA modes are of great interest in the analysis of mechanical and optical properties and can provide nanoscale diagnostics with twice the temporal resolution because of their slow propagation velocities (approximately half that of the LA mode). To exploit the shear acoustic properties, several techniques based on broken lateral symmetry have been developed: tight focusing of pump beams within a spot size of 1$\sim$2 $\mu$m to reduce the AC directivity in the depth direction\cite{Rossignol}, the transient interference patterns of two pumping beams\cite{Nelson}, LA-to-TA mode conversion at the isotropic--anisotropic crystal interface\cite{Hurley}, and excitations on anisotropic crystal surfaces\cite{Matsuda,Wen}.

In addition, under an external reverse bias in GaN-based piezoelectric diodes, the amplitude of the LA mode propagating along the $c$ axis \cite{Kim} and the localized TA mode along the non-$c$ axis have been independently enhanced \cite{Chen}. In most previous efforts, however, the growth direction was considered to be the crucial factor determining the strength of the AC phonon modes generated. From the viewpoint of prospective device applications, it would be useful if an external bias could play a role in modifying the selection rules for generating and detecting AC phonons. We recently demonstrated such a switching scheme of TA phonons in a laterally biased piezoelectric diode \cite{Jeong}. The amplitude and switching time of the TA mode could be electrically modulated. The structure with the electrically manipulated symmetry further revealed spectral shifts of the observed phonon frequencies due to modulation of the mechanical and optical constants.

This work examines the relative roles of the electric potential distributions in switching and amplitude modulation of AC phonons and presents the corresponding analytical expressions.  For comparison, we measured the AC phonon properties of samples with or without an asymmetric potential distribution. The detailed theoretical procedures are presented to describe AC modulation, including amplification and switching, on the basis of the electrically modulated elastic and dielectric properties.    The paper is organized as follows. We introduce the growth parameters, band structures, and strain-induced structural distortions of two contrasting samples, one under an asymmetrical potential distribution and the other under a symmetrical potential distribution, in Sec. II. Our experimental schemes are briefly described in Sec. III. In Sec. IV, we present the theoretical basis for electrical modulation of phononic properties, which is further classified into elastic tensor modulation for phonon generation and dielectric tensor modulation for optical detection of phonons. In Sec. V, we use our theoretical basis to explain the experimentally observed phononic spectra as a function of the external bias and dynamics. In Sec. VI, the role of crystal symmetry in modal detection is discussed in terms of the birefringence and detection sensitivity modulation under different potential distributions. Finally, we summarize our findings and conclusions in Sec. VII.

\section{Samples with different electric potential distributions}

\begin{figure}[!t]
\centering
\includegraphics[scale=0.22]{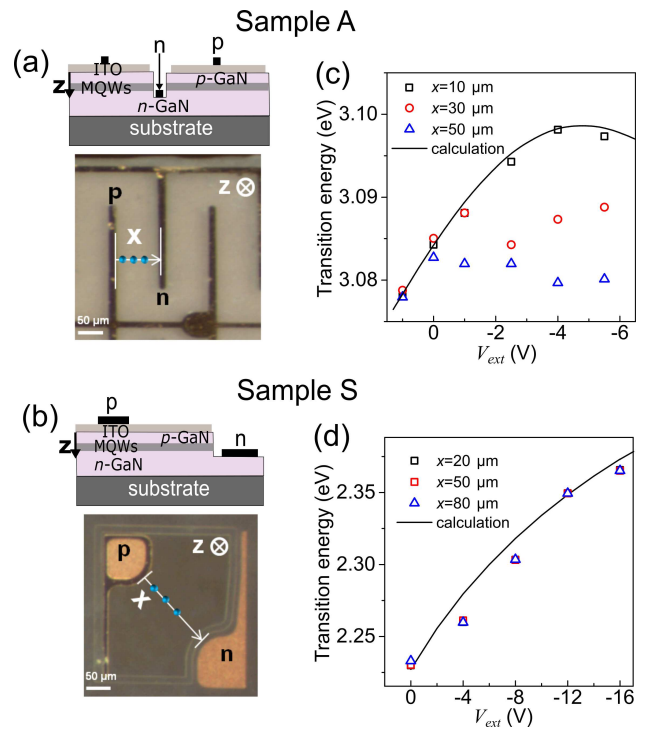}
\caption{
Side and top views of (a) sample A and (b) sample S, where the blue dots along the $x$ axis from the $p$ contact to the $n$ contact denote the laser excitation spots for the PL measurements. PL peak energy as a function of external bias in (c) sample A and (d) sample S.}
\end{figure}

Figure 1(a,b) shows micrographs of two representative structures with distinct potential distributions. They have similar vertical structures with multiple quantum wells (QWs) placed within the intrinsic region of $p$-$i$-$n$ diodes grown along the $c$ axis, where the reverse applied bias compensates for the piezoelectric field\cite{Jho2001}; the fundamental difference between them lies in the symmetry of the potential distributions. The sample under an asymmetric potential distribution (sample A hereafter), shown in Fig. 1(a), has a so-called interdigitated structure with narrow $p$ and $n$ electrodes that are laterally spaced by $\sim$100 $\mu$m along the $x$ direction on top of a thin indium tin oxide (ITO) layer ($\sim$40 nm). In this way, an electric current $I$ along the $x$ axis was unevenly applied in sample A. In contrast, the contact electrodes of the sample under a symmetric potential distribution (sample S hereafter), shown in Fig. 1(b), were much wider, with a thicker ITO layer ($\geq$150 nm) to evenly spread $I$ in the $x$--$y$ plane. The following layers were sequentially grown on sapphire substrates by metalorganic chemical vapor deposition in sample A (sample S): 3- (0.5-) $\mu$m-thick undoped GaN, 2.5- (3-) $\mu$m-thick $n$-GaN, six (five) QW layers of 2-nm-thick In$_{0.1}$Ga$_{0.9}$N (1.5-nm-thick In$_{0.25}$Ga$_{0.75}$N) encased by seven (six) 8- (7.5-) nm-thick GaN barriers, 120- (70-) nm-thick $p$-Al$_{0.05}$Ga$_{0.95}$N ($p$-Al$_{0.15}$Ga$_{0.85}$N), and 250- (140-) nm-thick $p$-GaN. The electron and hole concentrations in sample A (sample S) were estimated to be about 7 $\times$ 10$^{17}$ (2 $\times$ 10$^{17}$) cm$^{-3}$ and 8 $\times$ 10$^{17}$ cm$^{-3}$, respectively.

\begin{figure*}[!t]
\centering
\includegraphics[scale=0.8]{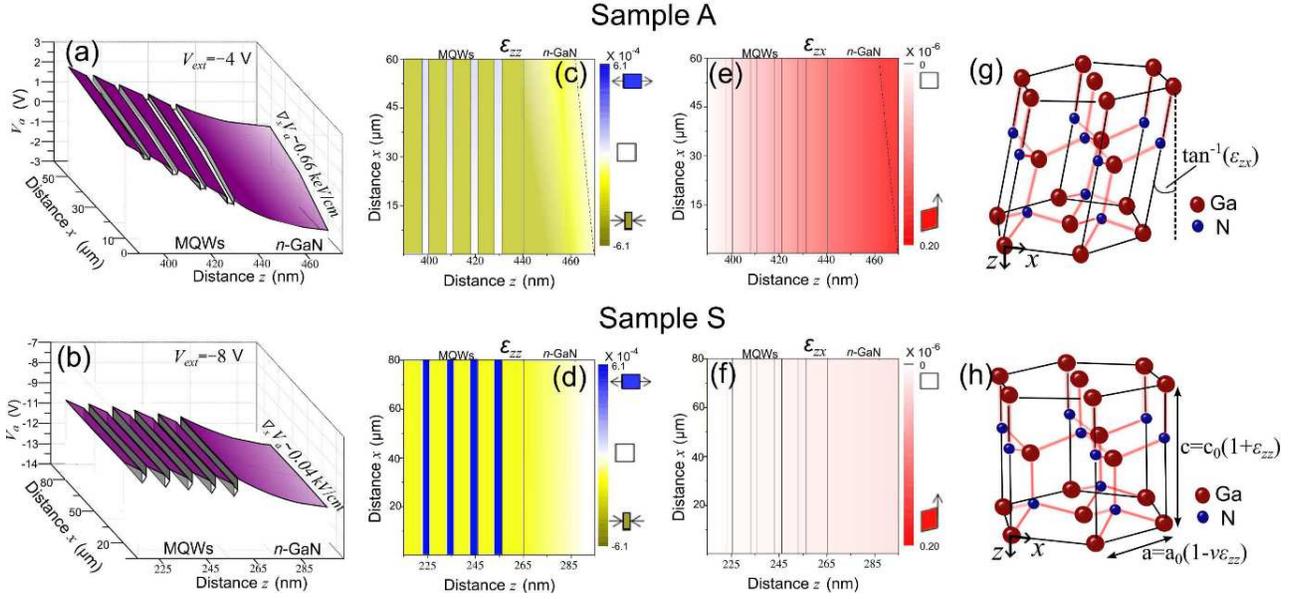}
\caption{
Electric potential profiles of (a) sample A and (b) sample S near MQWs/$n$--GaN interface. Distributions of normal strain $\varepsilon_{zz}$ in (c) sample A and (d) sample S. Shear strain $\varepsilon_{zx}$ distributions in (e) sample A and (f) sample S. Distortions in the lattice structures of (g) sample A and (h) sample S.}
\end{figure*}
%PL measurement and resultant energy band diagram

Figure 1(c,d) shows the results of spatially resolved photoluminescence (PL) measurements, where a frequency-doubled Ti:sapphire laser was used as an excitation source at 367.5 nm with a spot size of $\sim$10 $\mu$m. The emission properties of the samples differed greatly. As the excitation spot moved away from the $p$ electrodes, the PL peak energy was redshifted under the external bias $V_{ext}$ only in sample A. In the conventional case of sample S with an even current distribution, on the other hand, no notable energy shift was observed with $x$, as expected. The PL energy shift is well understood in terms of the quantum-confined stark effect \cite{Jho2002}. A typical PL peak for spontaneous emission should be matched with the lowest-lying QW transition and determined on the basis of the net electric field in the QW, $E_z^w$. Hence, the redshift of the PL peak energy with $x$ in sample A intuitively indicates that the actual magnitude of the vertically applied voltage decreased along $x$. To determine the energy band profiles in the $x-z$ plane, $E_z^w$ was extracted from the variational calculations of the PL peak energy. As $E_z^w$ and the net electric field in the barrier, $E_z^b$, are formulated using the built-in potential $V_{bi}$, the vertically applied voltage $V_a$ could be subsequently evaluated as a function of $x$ and $z$. The piezoelectric field $E_F$ was set to 0.98 MV/cm in sample A and 3.7 MV/cm in sample S, as indicated by the solid lines in Fig. 1(c,d).

The resultant electric potential profiles, incorporating the exerted potential distributions of $V_a(x,z)$, are shown in Fig. 2(a,b) for sample A at $-$4 V and sample S at $-$8 V, respectively. Following the procedures introduced above, the magnitude of $E_z^b$ ($\sim$0.8 MV/cm in sample A and $\sim$0.5 MV/cm in sample S) reproduced the rapid potential variations along the $z$ direction in both samples.  Further, the PL energy variations in Fig. 1(c) ultimately substantiated the increasing lateral electric field $E_{x} = -\nabla_{x} V_a$ in sample A, reaching $\sim$0.66 kV/cm at $-$4 V at the end of the depletion region. The magnitude of $E_{x}$ at $-$4 V was similar to that of $E_z^w$ in sample A but negligible in sample S. Accordingly, the tensor components of the mechanical strain are subject to distinct potential profiles in the piezoelectric structures exposed to the external bias. In particular, the additive lateral electric field $E_x$ in sample A could induce shear strain, which is otherwise forbidden.

According to the piezoelectric matrix, the strain components are connected by the electric field distributions:

\begin{equation}
\begin{bmatrix}
\varepsilon_{xx}\\
\varepsilon_{yy}\\
\varepsilon_{zz}\\
\varepsilon_{yz}\\
\varepsilon_{zx}\\
\varepsilon_{xy}\\
\end{bmatrix}
=
\begin{bmatrix}
0&0&d_{31}\\
0&0&d_{32}\\
0&0&d_{33}\\
0&d_{24}&0\\
d_{15}&0&0\\
0&0&0\\
\end{bmatrix}
\begin{bmatrix}
E_x\\
E_y\\
E_z\\
\end{bmatrix},
\end{equation}
where the piezoelectric strain constants are $d_{31}$ = $d_{32}$ = $-$1.9 pmV$^{-1}$, $d_{33}$ = 3.7 pmV$^{-1}$, and $d_{15}$ = $d_{24}$ = $-$3.1 pmV$^{-1}$ according to the previously published data for wurtzite GaN with $6mm$ symmetry\cite{Guy}. Then, the normal ($\varepsilon_{zz}$) and shear strain components ($\varepsilon_{zx}$) in the $x-z$ plane were calculated from the results in Fig. 1(c) [Fig. 1(d)] at $-$4 V for sample A (at $-$8 V for sample S) and are spatially mapped in Fig. 2(c--f), respectively. The signs of the normal strains ($\varepsilon_{zz}$) in Fig. 2(c,d) are opposite to those of the in-plane strains ($\varepsilon_{xx}$ and $\varepsilon_{yy}$), revealing positive Poisson ratios; the normal strains were compressive in the barriers and tensile in the QWs. On the other hand, the shear component ($\varepsilon_{zx}$) of sample A, shown in Fig. 2(e), became prominent throughout the biased region and was maximized near the end of the $n$-depletion region, in contrast to the suppressed values in sample S, shown in Fig. 2(f), due to the absence of the lateral electric field. The strain-induced structural distortions are illustrated schematically in Fig. 2(g,h). The volumetric deformation in Fig. 2(h) for sample S, where the lattice constants are modified to $c=c_0(1+\varepsilon_{zz})$ and $a=a_0(1-\nu \varepsilon_{zz})$ from the strain-free values of $c_0$ and $a_0$, implies that the structure is still laterally isotropic under an external bias $V_{ext}$. In contrast, in sample A under a shear strain $\varepsilon_{zx}$, monoclinic distortion that is geometrically defined by an angle $\theta$ to tilt the $c$ axis \cite{Nye} breaks the optical and mechanical symmetry in the $c$ plane in Fig. 2(g).

\section{Experimental schemes for measuring phonon dynamics}

Time-resolved pump--probe (PP) measurement and terahertz time-domain spectroscopy (THz-TDS) were conducted for various values of $V_{ext}$ or the angular parameters ($\theta,\phi$) to identify the role of the electric potential distributions in affecting both the detection and generation of AC phonons. A pair of Ti-sapphire lasers, synchronized at 76 MHz and with a jitter ($\sim$113 fs) smaller than the pulse width ($\sim$250 fs), was used at either 367.5 nm (for THz-TDS) or 367.5--375 nm (for PP measurement) in the reflective geometry, as shown in Fig. 3(a). For THz-TDS, each laser could be used as either a pumping source for THz generation or a signaling source (fixed at 800 nm) for detection via a Si lens coupled with a photoconductive antenna; the pump and probe excitation wavelengths were degenerate for PP measurement. The fluence of the pump beam for both PP measurement and THz-TDS was $\sim$85 $\mu$J/cm$^{2}$, and the ratio to that of the probe beam was 8:1. The laser spot size was maintained within $\sim$20 $\mu$m for the measurements. The angular parameters of the measurements are further illustrated in Fig. 3(b). The incident angle $\theta$ of the pump beam from the surface normal was fixed at 45$^{\circ}$ (for THz-TDS) or zero (for PP measurement), whereas the $\theta$ value of the probe beam was set to less than 5$^{\circ}$ unless otherwise mentioned (cf. Fig. 11--13).

\begin{figure}[!t]
\centering
\includegraphics[scale=0.28]{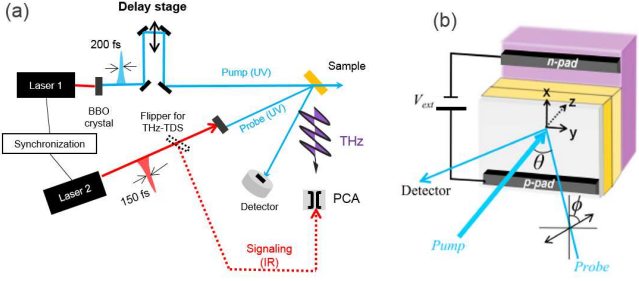}
\caption{
(a) Experimental scheme for time-resolved pump--probe and THz measurements using frequency-doubled Ti:sapphire lasers. (b) Illustration of measurement geometry for probing AC modal dynamics.}
\end{figure}

\section{Effects of electrically controlled crystal symmetry on AC phonon dynamics }

Applying an electric field parallel (perpendicular) to the 6-fold symmetry axis creates orthorhombic (monoclinic) distortion in a wurtzite structure via piezoelectricity. In conjunction with the types of structural deformations, the elastic and dielectric tensors are also classified into different symmetry groups. In this section, we develop mathematical formulations regarding the directional implications of the electric field components for modifying the acoustical properties of $c$-GaN.\\

\subsection{Electrical modulation of modal velocity}

We first develop mathematical formulations explaining changes in the crystal symmetry depending on the potential distributions and apply them to analytically express modal velocity variations. Under typical experimental conditions, the AC dynamics can be simply described by a loaded string model \cite{Chern,Stanton2005} assuming invariant crystal symmetry, that is, fixed elastic coefficients. In our work, the influence of symmetry variation on the elastic properties is revealed to explain the experimental results of electrically controlled AC modal generations and modulations in the propagation velocities, which are experimentally estimated in Sec. V.

As stated in Sec. II, the electric field components produce different types of strains. The effective elastic constants of a strained medium can be expressed as $C_{lm}=C_{lm}^0+C_{lmn}\varepsilon_{n}=C_{lm}^0+\Delta C_{lm}$, where $C_{lm}^0$ is the constant for the undeformed medium, and $C_{lmn}$ is the third-order elastic (TOE) tensor that creates the perturbation $\Delta C_{lm}$ under the applied strain $\varepsilon_{n}$. Such strain-induced elastic constant changes can alter the crystal symmetry \cite{Fuck}. For example, in $c$-GaN, the symmetry of $\Delta C_{lm}$ is still hexagonal for a normal strain $\varepsilon_{zz}$, whereas the symmetry becomes orthorhombic for in-plane normal strains ($\varepsilon_{xx}$ and $\varepsilon_{yy}$). When shear strains $\varepsilon_{zx}$ and $\varepsilon_{zy}$ are applied, the perturbed elastic tensors possess monoclinic symmetry with the diad axes parallel to the $y$ and $x$ directions, respectively. In our case, the vertical and lateral electric fields induced normal ($\varepsilon_{zz}=d_{33}E_z$) and shear strain ($\varepsilon_{zx}=d_{15}E_x$) via piezoelectricity. Therefore, the elastic tensor is expressed as
\begin{multline}
C_{lm}=C_{lm}^0+\Delta C_{lm}^H+\Delta C_{lm}^M \\
\simeq C_{lm}^0+C_{lm3}^H d_{33}E_z+C_{lm5}^M d_{15}E_x,
\end{multline}
%\small
%\begin{widetext}
%\begin{multline}
%C_{lm}=C_{lm}^0+\Delta C_{lm}^H+\Delta C_{lm}^M \\
%=
%\begin{bmatrix}
%C_{11}^0&C_{12}^0&C_{13}^0&0&0&0\\
%C_{12}^0&C_{11}^0&C_{13}^0&0&0&0\\
%C_{13}^0&C_{13}^0&C_{33}^0&0&0&0\\
%0&0&0&C_{44}^0&0&0\\
%0&0&0&0&C_{44}^0&0\\
%0&0&0&0&0&C_{66}^0\\
%\end{bmatrix}
%+
%\begin{bmatrix}
%C_{113}^H& C_{123}^H& C_{133}^H&0&0&0\\
%C_{123}^H& C_{113}^H& C_{133}^H&0&0&0\\
%C_{133}^H& C_{133}^H& C_{333}^H&0&0&0\\
%0&0&0& C_{443}^H&0&0\\
%0&0&0&0& C_{443}^H&0\\
%0&0&0&0&0& C_{663}^H\\
%\end{bmatrix}d_{33}E_z
%+
%\begin{bmatrix}
% 0& 0& 0&0&C_{155}^M &0 \\
% 0& 0& 0&0&C_{255}^M &0 \\
% 0& 0& 0&0&C_{355}^M &0 \\
%0&0&0&0&0&C_{465}^M\\
%C_{155}^M&C_{255}^M&C_{355}^M&0&0&0\\
%0&0&0&C_{465}^M&0&0\\
%\end{bmatrix}d_{15}E_x,
%\end{multline}
%\end{widetext}
where $\Delta C_{lm}^H$ ($\Delta C_{lm}^M$) is the perturbed elastic tensor due to vertical (lateral) electric fields, representing hexagonal (monoclinic) deformation, as illustrated in Fig. 2(h) [Fig. 2(g)]. The symmetry of the effective elastic tensor $C_{lm}$ consequently reveals that the structure could be monoclinically distorted \cite{Fuck, Nye} under asymmetric potential distributions. AC wave propagation in such structures is described by the Christoffel equation,
\begin{equation}
\nabla_{kl} C_{lm} \nabla_{mn} \mathrm{v}_n=\rho_0 \frac{\partial^2 \mathrm{v}_k}{\partial t^2},
\end{equation}
where $\mathrm{v}_k$ is the atomic displacement velocity along the $k$ axis, which determines the AC polarization. $\nabla_{kl}$ ($\nabla_{mn}$) is the divergence (gradient) matrix operator, which is given by

\begin{equation}
\nabla_{kl}=
\begin{bmatrix}
\frac{\partial}{\partial x}&0&0&0&\frac{\partial}{\partial z}&\frac{\partial}{\partial y}\\
0&\frac{\partial}{\partial y}&0&\frac{\partial}{\partial z}&0&\frac{\partial}{\partial x}\\
0&0&\frac{\partial}{\partial z}&\frac{\partial}{\partial y}&\frac{\partial}{\partial x}&0\\
\end{bmatrix},
\end{equation}

\begin{equation}
\nabla_{mn}=
\begin{bmatrix}
\frac{\partial}{\partial x}&0&0\\
0&\frac{\partial}{\partial y}&0\\
0&0&\frac{\partial}{\partial z}\\
0&\frac{\partial}{\partial z}&\frac{\partial}{\partial y}\\
\frac{\partial}{\partial z}&0&\frac{\partial}{\partial x}\\
\frac{\partial}{\partial y}&\frac{\partial}{\partial x}&0\\
\end{bmatrix}.
\end{equation}

On the basis of the modified elastic constants, the velocities of AC modes propagating along the $z$ axis with different polarizations are calculated to be
\small
%\begin{widetext}
\begin{multline}
\label{vla}
%v_{LA}=\frac{\sqrt{C_{33}+C_{44}+\sqrt{(C_{33}-C_{44})^2 +4C_{35}^2}}}{\sqrt{2\rho_0}}\\
%\simeq \frac{\sqrt{C_{33}+\frac{C_{35}^2}{C_{33}-C_{44}}}}{\sqrt{\rho_0}}
v_{LA}%\simeq \frac{\sqrt{C_{33}+\frac{C_{35}^2}{C_{33}-C_{44}}}}{\sqrt{\rho_0}}
\simeq \frac{\sqrt{C_{33}^0+C_{333}^H d_{33}E_z+\frac{(C_{355}^M)^2 }{C_{33}^0-C_{44}^0} d_{15}^2 E_x^2}}{\sqrt{\rho_0}},
\end{multline}
%\end{widetext}
\normalsize

\small
%\begin{widetext}
\begin{multline}
\label{vta}
%v_{TA\parallel x}=\frac{\sqrt{C_{33}+C_{44}-\sqrt{(C_{33}-C_{44})^2 +4C_{35}^2}}}{\sqrt{2\rho_0}}\\
%\simeq \frac{\sqrt{C_{44}-\frac{C_{35}^2}{C_{33}-C_{44}}}}{\sqrt{\rho_0}}\\
%\simeq \frac{\sqrt{C_{44}^0+C_{443}^H d_{33}E_z-\frac{(C_{355}^M)^2 }{C_{33}^0-C_{44}^0} d_{15}^2 E_x^2}}{\sqrt{\rho_0}},
v_{TA\parallel x}%=\frac{\sqrt{C_{33}+C_{44}-\sqrt{(C_{33}-C_{44})^2 +4C_{35}^2}}}{\sqrt{2\rho_0}}\\
%\simeq \frac{\sqrt{C_{44}-\frac{C_{35}^2}{C_{33}-C_{44}}}}{\sqrt{\rho_0}}\\
\simeq \frac{\sqrt{C_{44}^0+C_{443}^H d_{33}E_z-\frac{(C_{355}^M)^2 }{C_{33}^0-C_{44}^0} d_{15}^2 E_x^2}}{\sqrt{\rho_0}},
\end{multline} and
%\end{widetext}
\normalsize

\begin{equation}
v_{TA\parallel y}=\sqrt{\frac{C_{44}}{\rho_0}}= \sqrt{\frac{C_{44}^0 +C_{443}^H d_{33}E_z }{\rho_0}}.
\end{equation}

Equations (6--8) represent that the modal velocities are externally manipulatable by applying electric fields. We also note that, without $E_x$, the TA modal velocity with its polarization parallel to the $x$ axis ($v_{TA\parallel x}$) is the same as $v_{TA\parallel y}=\sqrt{\frac{C_{44}^0 +C_{443}^H d_{33}E_z}{\rho_0}} $. Interestingly, for $E_x$ (that is, for nonzero $\varepsilon_{zx}=d_{15}E_x $), $v_{TA\parallel x} < v_{TA\parallel y}$, indicating that the lateral degeneracy of TA phonon velocities is lifted.%broken. %We also note that not only $v_{TA}$ but also $v_{LA}$ are modified by $E_x$.

\subsection{Electrically controlled modal detection}

%which causes the amplitude of reflected probe beam to oscillate.
%In an isotropic medium, $\digamma^{iso}_LA$ is \cite{Stanton2005} for any probe polarization and $\digamma^{iso}_TA$ is zero for normal probe incidence due to lateral symmetry \cite{Stanton2005}.

In this section, we mathematically formulate the relationship between the electrically manipulated crystal symmetry and optical sensitivities of AC phonons. An AC wave packet $\eta_i(z,t)$ propagating in a material causes a spatiotemporal change in the refractive index $n$ by $\frac{\partial n}{\partial \eta_i}\eta_i(z,t)$, where $i$ corresponds to the mode of AC wavepackets.  In previous reports \cite{Stanton2005,Chern,Matsudadia,Matsudaoff}, Maxwell’s equations with phonon-induced refractive index modulations were developed for optically isotropic structures and used to obtain the differential reflectivity changes, $\Delta R/R$, in the probe pulse. Considering the dynamic Fabry--Perot (DFP) interference between the reflection from the surface, $r_0=\frac{1-n}{1+n}$, and the reflection from $\eta_i(z,t)$,  $\Delta r_i \propto \frac{\partial n}{\partial \eta_i} \eta_i(z,t)$, $\Delta R/R$ is simply expressed as $\int dz \digamma^{iso}_i \eta_i(z,t)$, \cite{Stanton2005} where $\digamma^{iso}_i$ is the detection sensitivity of $\eta_i(z,t)$ in an isotropic medium. In our experiments, however, the symmetry of the structures was externally controlled; hence, the detection mechanisms could differ from those previously reported.

Therefore, we first characterize the refractive index of an electrically distorted GaN by $E_x$ and $E_z$ by taking into account the symmetry-dependent photoelasticity. Then, the propagation of the probe light along the $z$ axis could be expressed by Maxwell's equation:
\begin{equation}
\{\begin{bmatrix}
-k_z^2& 0&0 \\
0& -k_z^2& 0 \\
0& 0&0 \\
\end{bmatrix}
+k^2(\mathbf{\epsilon^0}+\mathbf{\Delta \epsilon})\}
\begin{bmatrix}
\mathcal{E}^{\hbar \omega}_x\\
\mathcal{E}^{\hbar \omega}_y\\
0\\
\end{bmatrix}
=0,
\end{equation}
where $\mathcal{E}^{\hbar \omega}_x$ ($\mathcal{E}^{\hbar \omega}_y$) is the electric field intensity of the probe light along the $x$ ($y$) direction,  the wave number $k$ is $\omega_{probe}/c$, and $\bf{\epsilon^0}$ is the unperturbed dielectric tensor of wurtzite GaN. The strain-induced dielectric tensor perturbation, $\Delta \epsilon$, is described as
\small
\begin{widetext}
\begin{multline}
\mathbf{\Delta \epsilon}=
\begin{bmatrix}
p_{11}&p_{12}&p_{13}&0&p_{15}&0\\
p_{12}&p_{11}&p_{13}&0&p_{25}&0\\
p_{13}&p_{13}&p_{33}&0&p_{35}&0\\
0&0&0&p_{44}&0&p_{46}\\
p_{15}&p_{25}&p_{35}&0&p_{44}&0\\
0&0&0&p_{46}&0&p_{66}\\
\end{bmatrix}
\begin{bmatrix}
0\\
0\\
d_{33}E_z\\
0\\
d_{15}E_x\\
0\\
\end{bmatrix}
\\
=
\begin{bmatrix}
p_{13}d_{33}E_z+p_{15}d_{15}E_x\\
p_{13}d_{33}E_z+p_{25}d_{15}E_x\\
p_{33}d_{33}E_z+p_{35}d_{15}E_x\\
0\\
p_{35}d_{33}E_z+p_{44}d_{15}E_x\\
0\\
\end{bmatrix}
%\\
\equiv
\begin{bmatrix}
p_{13}d_{33}E_z+p_{15}d_{15}E_x& 0&p_{35}d_{33}E_z+p_{44}d_{15}E_x \\
0& p_{13}d_{33}E_z+p_{25}d_{15}E_x& 0 \\
p_{35}d_{33}E_z+p_{44}d_{15}E_x& 0&p_{33}d_{33}E_z+p_{35}d_{15}E_x \\
\end{bmatrix},
\end{multline}
\end{widetext}
\normalsize
where the symmetry of the photoelastic tensor $p_{lm}$ was determined by the structural distortion as  $p_{lm}=p_{lm}^0+\Delta p_{lm}=p_{lm}^0+p_{lmn}\varepsilon_{n}$ based on the unperturbed photoelastic constant $p_{lm}^0$. The detailed form of $p_{lm}$ is given in Appendix A.
After solving the eigenvalue problem in Eq. (9), the refractive indices were calculated to be

\small
\begin{widetext}
\begin{multline}
n_{\mathcal{E}^{\hbar \omega}_x}=
\frac{1}{\sqrt{\epsilon_e + p_{33}d_{33}E_z +p_{35}d_{15}E_x }}[\epsilon_o(\epsilon_e + p_{33}d_{33}E_z + p_{35}d_{15}E_x)+\epsilon_e( p_{13}d_{33}E_z+ p_{15}d_{15}E_x )\\
+d_{33}E_z d_{15}E_x(p_{15} p_{33}+ p_{35}p_{13}-2 p_{35}p_{44})+d_{33}^2E_z^2(p_{13} p_{33}- p_{35}^2)+d_{15}^2E_x^2( p_{15} p_{35}- p_{44}^2)]^{1/2}
\end{multline}
\end{widetext}
\normalsize
and
\begin{equation}
n_{\mathcal{E}^{\hbar \omega}_y}=\sqrt{\epsilon_o + p_{13}d_{33}E_z + p_{25}d_{15}E_x},
\end{equation}
where $\epsilon_o$ and $\epsilon_e$ are the ordinary and extraordinary components of $\bf{\epsilon^0}$, respectively. Similar to the case of AC waves, the refractive indices felt by the probe beam parallel to either the $x$ axis ($n_{\mathcal{E}^{\hbar \omega}_x}$) or the $y$ axis ($n_{\mathcal{E}^{\hbar \omega}_y}$) are the same without lateral electric fields; i.e., $n_{\mathcal{E}^{\hbar \omega}_x}$ =$n_{\mathcal{E}^{\hbar \omega}_y}=\sqrt{\epsilon_o + p_{13}d_{33}E_z}$. In the particular case with only a lateral field but without vertical fields, $n_{\mathcal{E}^{\hbar \omega}_x} \simeq \sqrt{\epsilon_o + p_{15} d_{15}E_x+d_{15}^2E_x^2(p_{15}p_{35}-p_{55}^2)/\epsilon_e}$ and $n_{\mathcal{E}^{\hbar \omega}_y} = \sqrt{\epsilon_o + p_{25}d_{15}E_x}$ become nondegenerate, demonstrating the lateral anisotropy of the refractive index.

The lateral anisotropy also affects the detection sensitivity $\digamma^{ani}_i$ for normally incident probe light as
\begin{equation}
\digamma^{ani}_{LA} \propto \frac{\partial n_{\mathcal{E}^{\hbar \omega}_x}}{\partial \eta_{3}}=\frac{\partial n_{\mathcal{E}^{\hbar \omega}_y}}{\partial \eta_{3}} \sim \frac{p_{13}}{2 \sqrt{\epsilon_o}},
\end{equation}
\begin{equation}
\digamma^{ani}_{TA\parallel x} \propto \frac{\partial n_{\mathcal{E}^{\hbar \omega}_x}}{\partial \eta_{5}} \sim \frac{1}{2 \sqrt{\epsilon_o}} (p_{15}+2d_{15}E_x(p_{15}p_{35}-p_{55}^2)/\epsilon_e),
\end{equation}
and
\begin{equation}
\digamma^{ani}_{TA \parallel y}\propto \frac{\partial n_{\mathcal{E}^{\hbar \omega}_y}}{\partial \eta_{5}} \sim \frac{p_{25}}{2 \sqrt{\epsilon_o}},
\end{equation}
where $\eta_{3}$ ($\eta_{5}$) is the strain of the LA ($x$-polarized TA) mode. As a result, the amplitude of $\Delta R/R$ for the $x$-polarized TA mode depends on the probe polarization $\phi$ due to the distinct formulations between $\digamma^{ani}_{TA\parallel x}$ and  $\digamma^{ani}_{TA\parallel y} $, where $\digamma^{ani}_{TA \parallel j}$ indicates $\digamma^{ani}_{TA}$ with its probe polarization parallel to the $j$ axis.

The electrically attained optical anisotropy is further characterized in consideration of the distinguished dependence of $\digamma_{i}$ on the probe incidence angle $\theta$.
When AC modes propagate in an isotropic medium along the $z$ axis, LA strain waves $\eta_{3}$ and TA strain waves $\eta_{5}$ locally modulate the dielectric tensor via the unperturbed photoelastic constants as
\small
\begin{multline}
\mathbf{\Delta \epsilon}
%\begin{bmatrix}
%p_{11}^0&p_{12}^0&p_{13}^0&0&0&0\\
%p_{12}&^0p_{11}^0&p_{13}^0&0&0&0\\
%p_{13}^0&p_{13}^0&p_{33}^0&0&0&0\\
%0&0&0&p_{44}^0&0&0\\
%0&0&0&0&p_{44}^0&0\\
%0&0&0&0&0&p_{66}^0\\
%\end{bmatrix}
%\begin{bmatrix}
%0\\
%0\\
%\eta_{3}\\
%0\\
%\eta_{5}\\
%0\\
%\end{bmatrix}\\
=
\begin{bmatrix}
p_{13}^0\eta_{3}\\
p_{13}^0\eta_{3}\\
p_{33}^0\eta_{3}\\
0\\
p_{44}^0\eta_{5}\\
0\\
\end{bmatrix}
\equiv
\begin{bmatrix}
p_{13}^0\eta_{3}& 0&p_{44}^0\eta_{5}\\
0& p_{13}^0\eta_{3}& 0 \\
p_{44}^0\eta_{5}& 0&p_{33}^0\eta_{3}\\
\end{bmatrix},
\end{multline}
\normalsize
where the diagonal and off-diagonal components of $\mathbf{\Delta \epsilon}$ are separated by $\eta_{3}$ and $\eta_{5}$. Owing to the modally separated tensor components, different modal sensitivities depending on $\theta$ are expected as
$\digamma^{iso}_{LA} \propto 4k'(1-\epsilon^{0})^{-1}p_{13}^0$ \cite{Matsudadia} and $\digamma^{iso}_{TA} \propto 2k_{x}k'k(k'-k'')^{-1}(\epsilon^{0} k'k''+k''^{2})^{-1}p_{44}^0$ \cite{Matsudaoff}, where $k=\omega_{probe}/c$, $k_{x}=k\sin(\theta)$, $k'=k\cos(\theta)$, and $k''=\sqrt{\epsilon^{0} k^{2}-k_{x}^{2}}$. Conversely, the contribution of the diagonal (off-diagonal) components to $\Delta R/R$ can be represented as $\digamma^{iso}_{LA}/p_{13}^0$ ($\digamma^{iso}_{TA}/p_{44}^0$). This explains why TA modes in an isotropic material cannot be detected with normal probe incidence, at which $\digamma^{iso}_{TA}$ becomes zero.

In contrast, in an anisotropic medium, the dielectric tensor is modulated by combinations of $\eta_{3}$ and $\eta_{5}$ via the perturbed photoelastic constants as
\small
\begin{multline}
\mathbf{\Delta \epsilon}=
%\begin{bmatrix}
%p_{11}&p_{12}&p_{13}&0&p_{15}&0\\
%p_{12}&p_{11}&p_{13}&0&p_{25}&0\\
%p_{13}&p_{13}&p_{33}&0&p_{35}&0\\
%0&0&0&p_{44}&0&p_{46}\\
%p_{15}&p_{25}&p_{35}&0&p_{44}&0\\
%0&0&0&p_{46}&0&p_{66}\\
%\end{bmatrix}
%\begin{bmatrix}
%0\\
%0\\
%\eta_{3}\\
%0\\
%\eta_{5}\\
%0\\
%\end{bmatrix}
%=
\begin{bmatrix}
p_{13}\eta_{3}+p_{15}\eta_{5}\\
p_{13}\eta_{3}+p_{25}\eta_{5}\\
p_{33}\eta_{3}+p_{35}\eta_{5}\\
0\\
p_{35}\eta_{3}+p_{44}\eta_{5}\\
0\\
\end{bmatrix}\\
\equiv
\begin{bmatrix}
p_{13}\eta_{3}+p_{15}\eta_{5}& 0&p_{35}\eta_{3}+p_{44}\eta_{5}\\
0& p_{13}\eta_{3}+p_{25}\eta_{5}& 0 \\
p_{35}\eta_{3}+p_{44}\eta_{5}& 0&p_{33}\eta_{3}+p_{35}\eta_{5}\\
\end{bmatrix}.
\end{multline}
\normalsize
Accordingly, measurements of the modal propagation are influenced by both diagonal and off-diagonal perturbations as $ \digamma^{ani}_{LA}=\alpha (\digamma^{iso}_{LA}/p_{13}^0) p_{13} + \beta (\digamma^{iso}_{TA}/p_{44}^0)p_{35}$ and $ \digamma^{ani}_{TA}=\alpha (\digamma^{iso}_{LA}/p_{13}^0) p_{15} + \beta (\digamma^{iso}_{TA}/p_{44}^0)p_{44}$. $\alpha$ ($\beta$) is an empirical factor representing the mixed contribution of the diagonal (off-diagonal) perturbation in $\mathbf{\Delta \epsilon}$ to the signal amplitude of mode $i$ via, e.g., the modified LA (TA) sensitivity due to the interaction between TA (LA) waves and probe beams. The sensitivities in the anisotropic region are now simplified in terms of those in the isotropic region as

\begin{multline}
\begin{pmatrix}
\digamma_{LA}^{ani}\\
\digamma_{TA}^{ani}\\
\end{pmatrix}
=
\begin{pmatrix}
\alpha p_{13}/p_{13}^0&\beta p_{35}/p_{44}^0\\
\alpha p_{15}/p_{13}^0&\beta p_{44}/p_{44}^0\\
\end{pmatrix}
\begin{pmatrix}
\digamma_{LA}^{iso}\\
\digamma_{TA}^{iso}\\
\end{pmatrix},
%\\
%=
%\begin{pmatrix}
%\alpha(1+\Delta p_{13}/p_{13}^0)&\beta\Delta p_{35}/p_{44}^0\\
%\alpha\Delta p_{15}/p_{13}^0&\beta(1+\Delta p_{44}/p_{44}^0)\\
%\end{pmatrix}
%\begin{pmatrix}
%\digamma_{LA}^{iso}\\
%\digamma_{TA}^{iso}\\
%\end{pmatrix},
\end{multline}
and are further compared with the experimental results in Sec. VI.

\subsection{Acoustic screening effect on modal sensitivity}
 \begin{figure}[!t]
\centering
\includegraphics[scale=0.44]{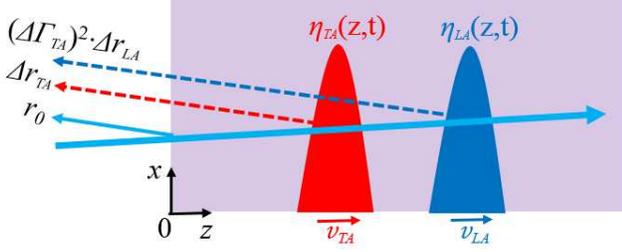}
\caption{Illustration of the $\Delta R/R$ measurement scheme for two different AC wave packets, $\eta_{LA}(z,t)$ and $\eta_{TA}(z,t)$.}
\end{figure}

An AC wave packet locally perturbs the refractive index of a material, inducing modulations of the light reflection, $\Delta r$, and a corresponding transmission change, $\Delta \Gamma=-\Delta r$, assuming that the phonon-induced absorption change is negligible. In most ultrasonic studies with a single AC mode, these modulations, $\Delta r$ and $\Delta \Gamma$, have been investigated independently using different PP geometries for reflectivity and transmission.  However, for measurements of more than one AC mode, the combined effect of the modulations on the detection sensitivity should be taken into account, since the $\Delta r$ ($\Delta \Gamma$) value of a mode can influence the $\Delta \Gamma$ ($\Delta r$) value of other modes.

Figure 4 shows the typical propagation of the strain pulses, the faster $\eta_{LA}(z,t)$ and the slower $\eta_{TA}(z,t)$, induced by the LA and TA mode in sample A, where the modal amplitudes are electrically controlled (cf. Fig. 5). The propagation of $\eta_{TA}(z,t)$, which lags behind $\eta_{LA}(z,t)$ because its propagation velocity is $\sim$50$\%$ slower, can be simply traced by measuring the DFP interference between the reflection from the surface, $r_0$, and the reflection from $\eta_{TA}(z,t)$, $\Delta r_{TA}$ \cite{Stanton2005}. However, the light reflection from $\eta_{LA}(z,t)$ is additionally modulated by a factor of $(\Delta \Gamma_{TA})^2$, as $\eta_{TA}(z,t)$ screens the probe beam incident on and reflected from $\eta_{LA}(z,t)$, which modifies the sensitivity function $\digamma_{LA}$ and $\Delta R/R$ for the LA mode by the same amount. This acoustic screening effect on the modal sensitivity is experimentally verified in Sec. VI.

\section{Role of symmetry in the acoustic phonon generation}

In this section, we show that the application of an external bias in the \textit{c} plane can break the selection rules of isotropic $c$-GaN, switching on the normally forbidden TA mode. To explicitly show the differing roles of vertical and lateral electric fields in the generation process, a strain-dependent model of the AC wave equation was developed. Our experimental results further demonstrate that the modal propagation velocities are controllable under an external bias. In addition, the switching time of the TA mode can be effectively manipulated by changing the scale of the laterally biased region.

\subsection{Manipulation of AC mode selection rule}

\subsubsection{Strain-based description of AC wave generation}
 When coherent AC phonons are displacively initiated via the transient electric field screening by the photocarriers, the relevant dynamic features can be described on the basis of a typical loaded-string equation \cite{Chern},

%, $C_{i}$ is the elastic constant ($C_{33}$ for $C_{LA}$ and $C_{44}$ for $C_{TA}$, experimentally estimated as $C_{LA}$=344 GPa and $C_{TA}$=129 GPa), and $\rho_{0}$ is the mass density.
\begin{equation}
\frac{{\partial}^{2}u_i}{{\partial}t^{2}}-v_i^2 \frac{{\partial}^{2}u_i}{{\partial}z^{2}}
=S_i(x,z,t),
\end{equation}
where $i$ stands for each phonon mode (LA or TA), $u_{i}$ represents the atomic displacement, and $v_i$ is the modal velocity formulated in Sec. IV. The general form of driving force $S_{i}(x,z,t)$ is expressed as

\begin{equation}
\label{eq:S-sigma}
\begin{bmatrix}
S_{xx}\\
S_{yy}\\
S_{zz}\\
S_{zy}\\
S_{zx}\\
S_{xy}\\
\end{bmatrix}
=
\begin{bmatrix}
\Delta \sigma_{xx}\\
\Delta \sigma_{yy}\\
\Delta \sigma_{zz}\\
\Delta \sigma_{zy}\\
\Delta \sigma_{zx}\\
\Delta \sigma_{xy}\\
\end{bmatrix}
=
\begin{bmatrix}
0&0&e_{31}\\
0&0&e_{31}\\
0&0&e_{33}\\
0&e_{24}&0\\
e_{15}&0&0\\
0&0&0\\
\end{bmatrix}
\begin{bmatrix}
\Delta E_x\\
\Delta E_y\\
\Delta E_z\\
\end{bmatrix},
\end{equation}
where $\mathbf{\Delta \sigma}$ is the instantaneous stress change, $e_{ij}$ is the piezoelectric stress constant, and $\Delta E_j$ is the screened electric field intensity along the direction $j$. The driving force can also be represented in terms of the instantaneous strain change $\mathbf{\Delta \varepsilon}$ as

\small
%\begin{widetext}
\begin{multline}
\label{eq:sigma-epsilon}
\begin{bmatrix}
\Delta \sigma_{xx}\\
\Delta \sigma_{yy}\\
\Delta \sigma_{zz}\\
\Delta \sigma_{zy}\\
\Delta \sigma_{zx}\\
\Delta \sigma_{xy}\\
\end{bmatrix}
\simeq
[C^0_{lm}][d_{ij}]
%\begin{bmatrix}
%C_{11}&C_{12}&C_{13}&0&0&0\\
%C_{12}&C_{11}&C_{13}&0&0&0\\
%C_{13}&C_{13}&C_{33}&0&0&0\\
%0&0&0&C_{44}&0&0\\
%0&0&0&0&C_{44}&0\\
%0&0&0&0&0&C_{66}\\
%\end{bmatrix}
%\begin{bmatrix}
%\Delta \varepsilon_{xx}\\
%\Delta \varepsilon_{yy}\\
%\Delta \varepsilon_{zz}\\
%\Delta \varepsilon_{zy}\\
%\Delta \varepsilon_{zx}\\
%\Delta \varepsilon_{xy}\\
%\end{bmatrix}
%\\
%=
%\begin{bmatrix}
%C_{11}&C_{12}&C_{13}&0&0&0\\
%C_{12}&C_{11}&C_{13}&0&0&0\\
%C_{13}&C_{13}&C_{33}&0&0&0\\
%0&0&0&C_{44}&0&0\\
%0&0&0&0&C_{44}&0\\
%0&0&0&0&0&C_{66}\\
%\end{bmatrix}
%\begin{bmatrix}
%0&0&d_{31}\\
%0&0&d_{31}\\
%0&0&d_{33}\\
%0&d_{24}&0\\
%d_{15}&0&0\\
%0&0&0\\
%\end{bmatrix}
%\begin{bmatrix}
%\Delta E_x\\
%\Delta E_y\\
%\Delta E_z\\
%\end{bmatrix}
%\\
%=
%\begin{bmatrix}
%0&0&C_{11}d_{31}+C_{12}d_{31}+C_{13}d_{33}\\
%0&0&C_{12}d_{31}+C_{11}d_{31}+C_{13}d_{33}\\
%0&0&C_{13}d_{31}+C_{13}d_{31}+C_{33}d_{33}\\
%0&C_{44}d_{24}&0\\
%C_{44}d_{15}&0&0\\
%0&0&0\\
%\end{bmatrix}
\begin{bmatrix}
\Delta E_x\\
\Delta E_y\\
\Delta E_z\\
\end{bmatrix}
\\
\simeq
\begin{bmatrix}
0&0&C^0_{11}d_{31}+C^0_{12}d_{31}+C^0_{13}d_{33}\\
0&0&C^0_{12}d_{31}+C^0_{11}d_{31}+C^0_{13}d_{33}\\
0&0&C^0_{13}d_{31}+C^0_{13}d_{31}+C^0_{33}d_{33}\\
0&C^0_{44}d_{24}&0\\
C^0_{44}d_{15}&0&0\\
0&0&0\\
\end{bmatrix}
\begin{bmatrix}
\frac{\Delta \varepsilon_{zx}}{d_{15}}\\
\frac{\Delta \varepsilon_{zy}}{d_{24}}\\
\frac{\Delta \varepsilon_{zz}}{d_{33}}\\
\end{bmatrix},
\end{multline}
%\end{widetext}
\normalsize
where $[C^0_{lm}]$ is the unperturbed elastic tensor and $[d_{ij}]$ is the piezoelectric strain tensor. Therefore, $\Delta \sigma_{zz}$ for $S_{LA}$ (=$S_{zz}$) is $(2 \frac{d_{31}}{d_{33}}C^0_{13}+C^0_{33})\Delta \varepsilon_{zz}$, where $\frac{d_{31}}{d_{33}} \sim -0.5$ represents the Poisson effect, and $\Delta \sigma_{zx}$ for $S_{TA}$ (=$S_{zx}$) is $C^0_{44}\Delta \varepsilon_{zx}$. By assuming complete field screenings in the epicenters, which were confirmed by the saturated signal amplitudes in the fluence range shown in Fig. 6(b), the modal driving force is

\begin{equation}
S_i(x,z,t)=\frac{1}{\rho_{0}}\bar{C}_i\varepsilon_i(x,z)H(t),
\end{equation}
where the roles of the externally applied strains and photocarriers are clearly distinguished. The selection of the AC mode is determined by the strains, $\varepsilon_{LA}$ ($\varepsilon_{zz}$) and $\varepsilon_{TA}$ ($\varepsilon_{zx}$), with the corresponding effective modal constants, $\bar{C}_{LA}=2\frac{d_{31}}{d_{33}}C^0_{13}+C^0_{33}$ and $\bar{C}_{TA}=C^0_{44}$, respectively. The extra term $2\frac{d_{31}}{d_{33}}C^0_{13}$ in $\bar{C}_{LA}$ resulted from the contribution of the in-plane strain changes ($\Delta \varepsilon_{xx}$ and $\Delta \varepsilon_{yy}$) to the normal stress change $\Delta \sigma_{zz}$. The instantaneous initiation of each mode by abrupt field screening is described by the Heaviside step function $H$$(t)$ without regard to the mode index $i$.  In this description, only the piezoelectric effect was considered because of its dominant role in the generation process in samples grown along the $c$ axis \cite{Chern}.

\subsubsection{Electrically controlled AC mode generations}

\begin{figure}[!t]
\centering
\includegraphics[scale=0.13]{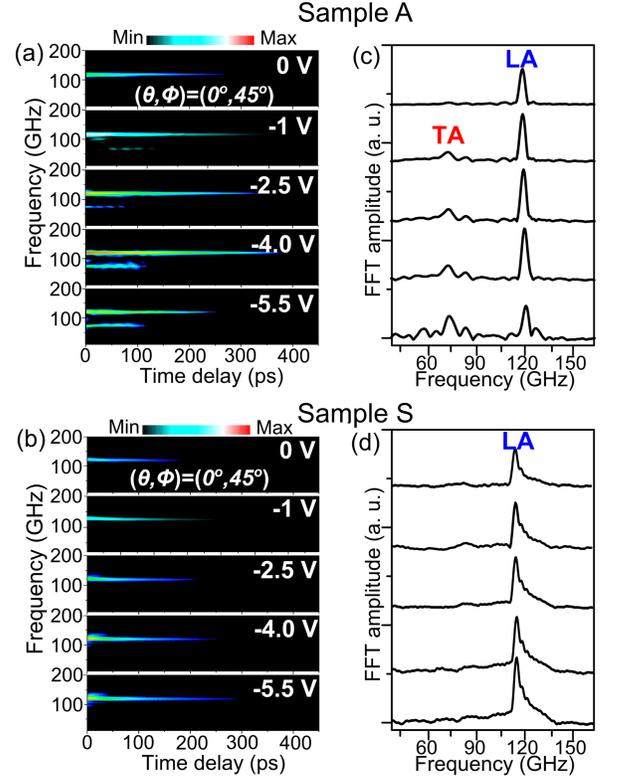}
\caption{SWFFTs of the DFP interference from (a) sample A and (b) sample S under reverse bias from 0 to $-$5.5 V. The contour plots show spectral information (vertical axis) with respect to the time delay (horizontal axis). Integrated spectra over the entire temporal range for (c) sample A and (d) sample S.}
\end{figure}

Ultrafast modification of the potential profiles by photocarriers initiates the propagating AC waves; the probe laser waves encounter the modal AC waves with different positions and with spatial extents and are partially reflected by or transmitted through the wave packets \cite{Yahng,Stanton2005}. Consequently, the multiple reflections of the probe laser waves form a DFP series, the oscillation frequency components of which are expressed as $f_{i}$=$2v_{i}n$cos($\theta_{t}$)/$\lambda$$_{probe}$, where $\theta_{t}$ is the angle of the probe transmission inside the materials and $\lambda$$_{probe}$ is the wavelength of the probe beam. The transient oscillations in reflective PP measurements under $V_{ext}$ have accordingly been extracted from $\Delta R/R$ based on the sliding-window fast Fourier transform (SWFFT) spectra, as shown in Fig. 5(a,b). In sample A, not only $f_{LA}$ at 119 GHz but also the novel spectral component at 70 GHz emerged under an external bias $V_{ext}$. $f_{LA}$ matches well the $v_{LA}$ value of $\sim$7300 m/s; this was further confirmed by the time of flight, which was revealed by an abrupt phase jump at the GaN-sapphire interface\cite{Yahng}, whereas the lower-frequency component (70 GHz) was ascribed to a new AC mode with $v_{TA}$ ($\sim$4200 m/s \cite{Yamaguchi}) and could not be explained by any previous reports in isotropic structures. In clear contrast, sample S without lateral fields showed a monochromatic spectral component the amplitude of which increased gradually with the reverse external bias from 0 to $-$5.5 V.

To verify that the 70 GHz component from sample A originated from the TA mode, the prominent frequencies measured with various values of  $\lambda_{probe}$ were converted into the corresponding velocities, as shown in Fig. 6(a). The external bias was held at $-$4 V throughout the experiment, whereas the configurations of the pump and probe energies were degenerate except for one nondegenerate case (3.32 and 1.66 eV, respectively) in Fig. 6(a). The high-frequency components (blue triangles) agreed well with the expected value of $v_{LA}$ over the entire range. The 70 GHz component (red squares) also conformed to the $\lambda$$_{probe}$-dependent DFP characteristics of $f_{TA}$ at constant $v_{TA}$.

In terms of the modal amplitudes, the bias-dependent changes in sample A were further explained by the different distributions of the normal and shear strains in Fig. 2(c,e), respectively. To investigate the generation mechanism and relevant scale, the AC modal components in the Fourier-transformed spectra were traced as a function of the pump fluence at $-$4 V with a fixed laser wavelength of 367.5 nm, as shown in Fig. 6(b). The amplitudes for both AC modes were saturated around 80 $\mu$J/cm$^{2}$, indicating that the generation mechanisms for both modes were dominated by transient field screening rather than deformation potential coupling \cite{Wen}.

The strength of the screening electric field and the width of the epicenters can be estimated using a simple virtual capacitor model \cite{CTYu} formed of the accumulated photocarrier distributions. For coherent LA phonon generation, electric fields transiently induced by photocarriers within multiple QWs (referred to as in-well screening \cite{Kim}) play a major role. As $V_{ext}$ increases, however, the reduced effective barrier widths result in considerable electronic tunneling, together with thermionic emission of holes over the lower valence band barriers \cite{Jho2002}. In this way, the charge densities of the serially connected virtual capacitors within the multiple QW region decrease; thus, the contribution of the QW carriers to AC mode generation also decreases. Therefore, the suppressed LA mode signal after $-$4 V in Fig. 5(a,c) implies that LA phonons originate from the QWs.

\begin{figure}[!t]
\centering
\includegraphics[scale=0.12]{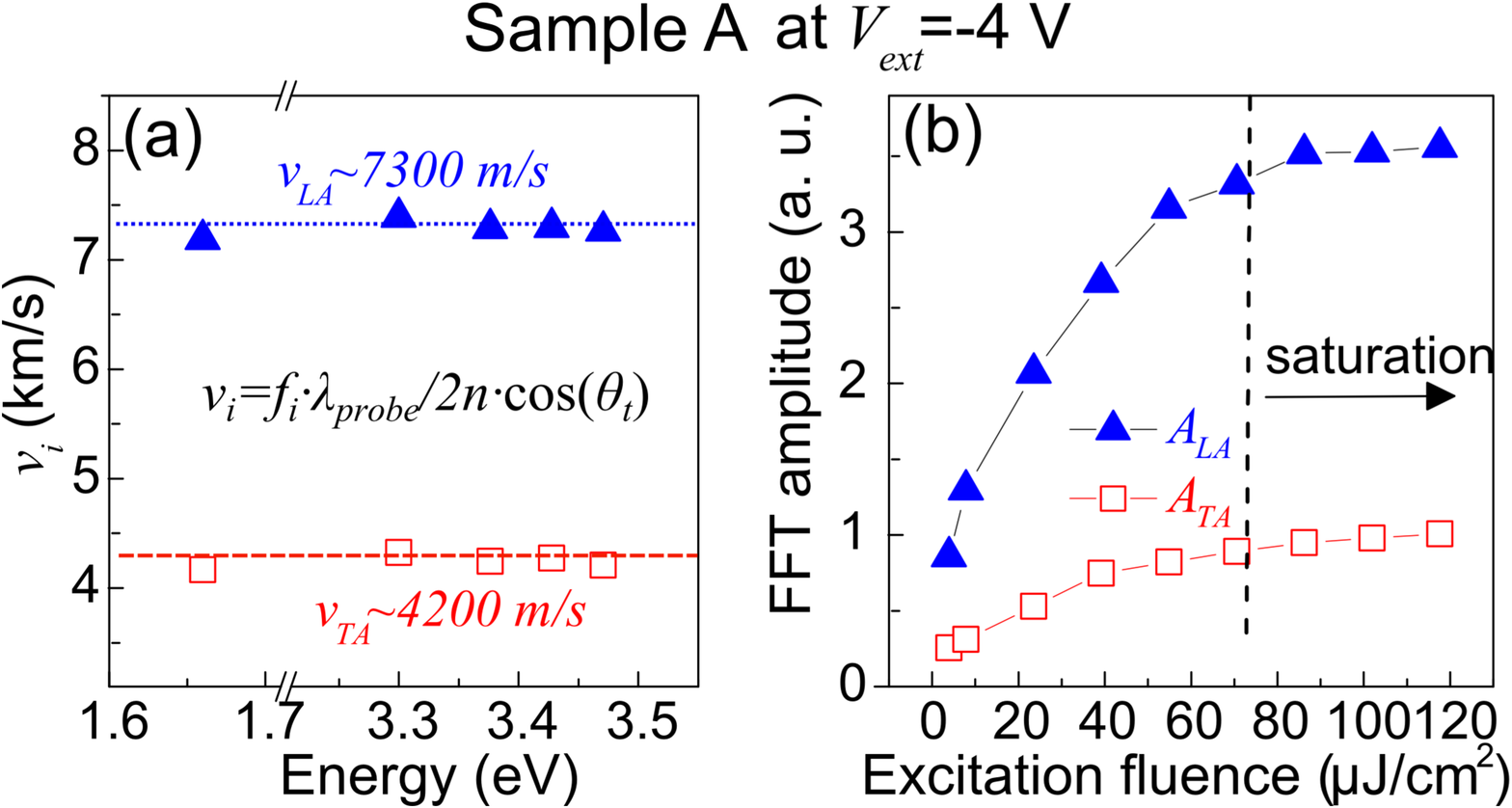} \caption{
(a) Modal velocities extracted from the DFP oscillations in sample A as a function of probe beam energy. (b) Spectral peak amplitudes of DFP oscillations as a function of pump fluence.}
\end{figure}

At a large external bias $V_{ext}$, the macroscopic screening, the spatial extent of which encompasses  the intrinsic and depletion regions (referred to as out-of-well screening \cite{Jahn}) may possibly become as important as the in-well screening. Under these conditions, the photocarrier sweep-out under the vertical electric field ($\sim$0.8 MV/cm at $-$4 V) leads to virtual capacitor formation, where the relevant sheet carrier densities are obtained by integrating the photocarrier distributions over the screening distance. Using the photocarrier density observed under a pump fluence of 80 $\mu$J/cm$^{2}$ and the high-frequency dielectric constant \cite{Barker}, the required screening distance to compensate for the electric field of $\sim$0.8 MV/cm was estimated to be $\sim$40 nm. Because of the significant saturation velocity difference between electrons and holes in GaN (a factor of $\sim$12 \cite{Wraback,Lin2002}), the photoexcited electrons immediately accumulate at the boundary of the $n$-depletion region within $\sim$0.4 ps, whereas the much slower holes take more than 5 ps to cross the entire macroscopic region. Therefore, during the AC mode generation period ($\sim$0.5 ps), a virtual capacitor scaled around 40 nm \cite{CTYu} could be spatially formed near the $n$-depletion region where shear strains are prominent, as shown in Fig. 2(e). In this regard, the TA signal, whose amplitude persistently grows with $V_{ext}$ despite QW tunneling, is thought to be generated from the $n$-depletion region.

\subsection{Symmetry-dependent AC mode dynamics}

\subsubsection{Mode-dependent temporal lineshape analysis}

%lineshape analysis of PP signals Band-pass-filtered signals under 4 V around the LA (108--144 GHz) and TA (52--88 GHz) modes

The propagation directions and spatial origins of the AC modes can be explicitly identified by tracing the amplitudes and phase changes of the oscillatory components of $\Delta R/R$. To explore different modal behaviors in sample A, the oscillatory signal (black line) at $-$4 V was band-pass-filtered at around the TA and LA modes (yellow scattered lines), as shown in Fig. 7(a). Because of the limited skin depth $\xi$ of the probe beam ($\sim$700 nm), the envelope for a descending AC wave packet ($\eta^{d}_{i}$, where $i$ indicates the phonon mode, LA or TA) would decay with time, whereas the envelope for an ascending wave packet ($\eta^{a}_{i}$) grows as $\eta^{a}_{i}$ approaches the surface \cite{Lin}. The amplitude change of each mode for $0<t<\tau_{i}$ in Fig. 7(a) could be fitted by considering both $\eta^{a}_{i}$ and $\eta^{d}_{i}$ components (red and blue lines).

%These spatially concentrated AC strains could be simplified into $\eta_{i}^{a}$=$\int_{D_i^{a}}\varepsilon_{i}(z)dz\cdot\delta(z+v_{i}t)$ for the ascending waves and $\eta_{i}^{d}$=$\int_{D_i^{d}}\varepsilon_{i}(z)dz\cdot\delta(z-v_{i}t)$ for the descending waves, where the region of integration $D_i^{d}$  ($D_i^{a}$) corresponds to the SDR (either to the MQW region for the LA mode or to the $n$-depletion region for the TA mode).

When the values of $\tau_{i}$ are multiplied by the corresponding velocities, they are converted into the propagation distances of the AC phonons ascending from the epicenters to the surface (and vice versa for $\eta^{d}_{i}$). The distance for the LA mode ($\sim$415 nm) coincides well with the distance from the center of the intrinsic region to the surface (405 nm). Therefore, $\eta^{a}_{LA}$ is considered to be generated mainly from the MQWs, as in conventional cases \cite{Yahng,Stanton2005}. In contrast, the TA mode traveled $\sim$80 nm farther than the LA mode, which is approximately the sum of half the intrinsic region and $n$-depletion region widths, $l_D(n_e,n_p,V_{ext})$ \cite{Jho2002}. Thus, $\eta^{a}_{TA}$ was considered to be launched mainly from the $n$-GaN depletion region, as can be further inferred from the shear strain distribution, which was maximized near the $n$-depletion region in Fig. 2(e). For this reason, the slightly increased value of $\tau_{TA}$ between $-$1 V and $-$5.5 V ($\sim$ 20 ps) could be converted into the enlarged travel distance of 84 nm in accordance with the expanded depletion region width at the surface and in the $n$-GaN region. This agreement between the transient digitized appearance and the travel path of the TA mode in Fig. 7(a) further confirmed that TA modes can be not only generated but also detected only in the laterally biased anisotropic region. This electrically manipulated AC modal selection also agreed with the shear strain distributions in the biased region in Fig. 2(e).

The initial deviations of the AC modes from the fitting curves are attributed to partial propagation of the surface-generated strain components toward the surface\cite{Thomsen} considering the relatively wide surface depletion region (SDR, $\sim$150 nm)\cite{Foussekis}. Without an ITO layer, the reflection of an AC mode from the air/$p$-GaN interface would be almost 100$\%$\cite{Lin, Thomsen} without diffuse phonon scattering. In our case, however, the AC mode reflection from the ITO/$p$-GaN interface is estimated to be $\sim$0.25 (0.19) for the LA (TA) mode, on the basis of the acoustic impedance mismatch between GaN\cite{Polian} and ITO\cite{Wittkowski}.

\begin{figure}[!t] \centering
\includegraphics[scale=0.13]{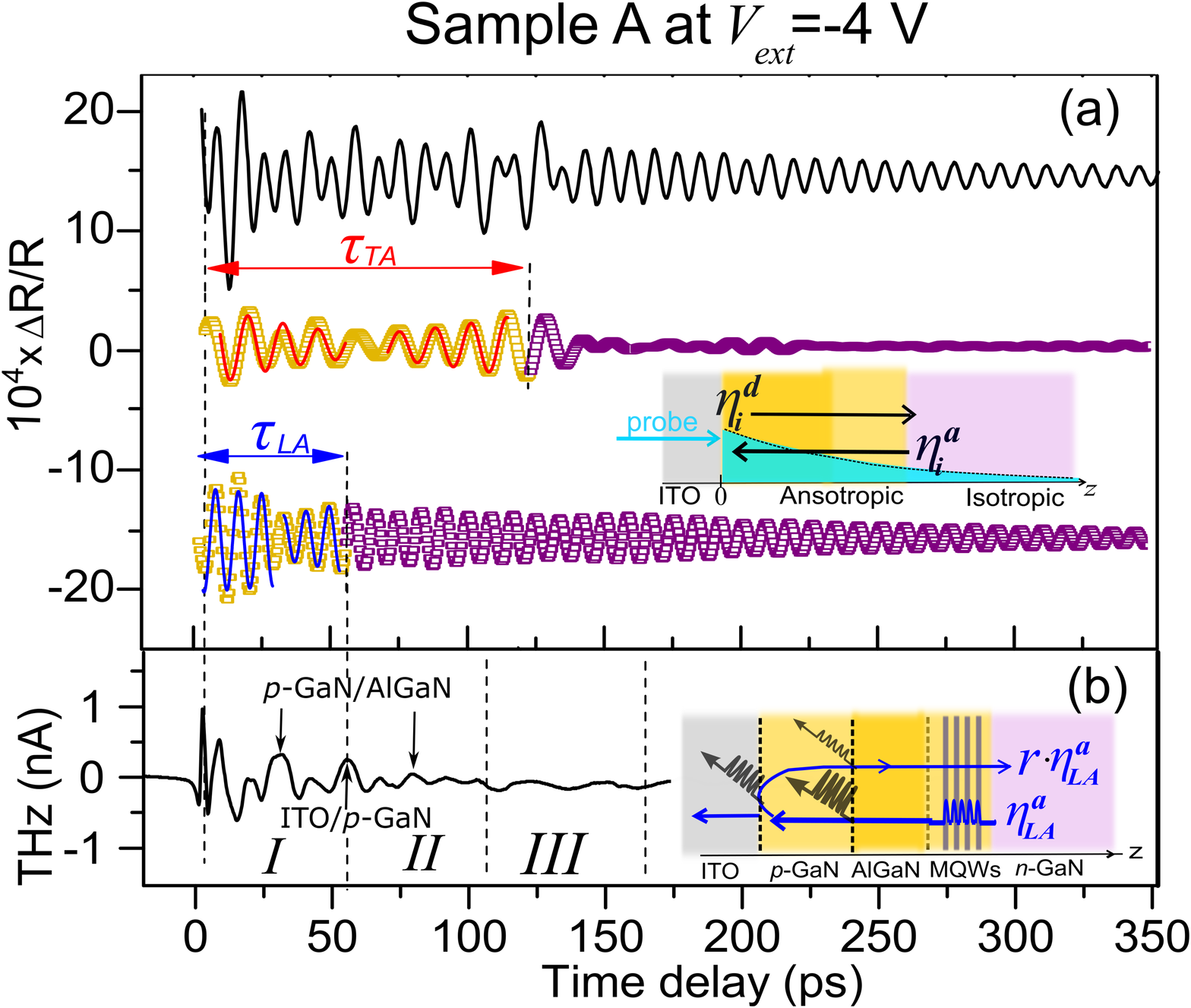} \caption{
(a) Total oscillatory component in PP signal (black line) was filtered into the TA (52--88 GHz) and LA mode (108--144 GHz) frequencies. Inset illustrates the spatial origins and propagations of the AC modes. (b) THz radiation as a function of time delay. Inset correlates the THz signal with the propagation of the ascending LA wave packet. }
\end{figure}

The AC reflectivity at the $p$-GaN/ITO interface, $r$, together with the spatial origins were further investigated by measuring the emitted THz waves during AC propagations in Fig. 7(b). When an AC wave packet $\eta_i$ traverses a piezoelectric interface, transient polarization currents are induced as $j_i^{int}\propto -(1/d_{1,i}-1/d_{2,i})v_{i} \eta_i (z_{int}-v_{i}t)$, where $d_{1,i}$ and $d_{2,i}$ are the piezoelectric strain constants of the adjacent layers for mode $i$, as previously reported for LA waves \cite{Armstrong,Stanton2003}. Therefore, the transient THz emission patterns directly reflect the AC wave profiles crossing the interfaces at the position of $z_{int}$ as illustrated in the inset of Fig. 7(b); e.g., the temporal width of a THz pulse coincides with the spatial width of $\eta_{i}$ divided by $v_{i}$.

To reduce complexity in the analysis, we focused on the propagation of the most intense strain component, $\eta^{a}_{LA}$, dividing the THz signal into three temporal domains. In time domain $I$ ($0<t<\tau_i$), $\eta^{a}_{LA}$ travels from the multiple QWs toward the ITO/$p$-GaN interface. The fast signal at the beginning is probably explained by surge photocurrent generation in the sample\cite{Mittleman}. The signal at $t\sim\tau_{LA}-25 \ ps$ (denoted by arrow) is attributed to the interaction between $\eta^{a}_{LA}$ and the $p-$GaN/AlGaN interface. In time domain $II$ ($\tau_{i}<t<2\tau_{i}$), the AC packet is reflected from the ITO/$p$-GaN interface with a reduced amplitude of $r\cdot\eta^{a}_{LA}$. The amplitude ratio ($\sim$30$\%$) of the signals at $t\sim\tau_{LA}\pm25 \ ps$ (denoted by arrows) was in good agreement with the reflection of $\sim$25$\%$ estimated from the AC impedance mismatch. The lingering slow oscillations in time domain $III$ ($2\tau_{i}<t<3\tau_{i}$) are suspected to be generated by AC phonon--carrier interactions in the $n$-GaN side \cite{Stanton2003}.

\subsubsection{Electric-field-induced propagation velocity modulation}

\begin{figure}[!t] \centering
\includegraphics[scale=0.19]{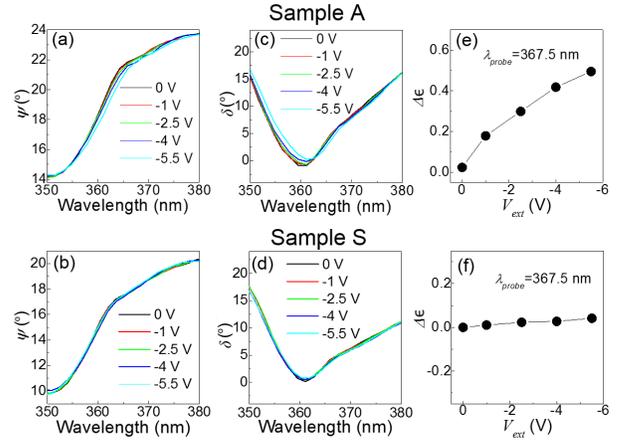} \caption{
Measurements of ellipsometric parameters (a,b) $\psi$ and (c,d) $\Delta$ from  sample A and sample S for different $V_{ext}$ at wavelengths ranging from 350 to 380 nm. $V_{ext}$-induced modulations in dielectric constants for (e) sample A and (f) sample S.}
\end{figure}

Another intriguing feature of the electrically controlled symmetry is the spectral blueshifts of $f_i$ in both samples under increasing $V_{ext}$ in Fig. 5(c,d). To investigate the external-field-dependent dynamic properties of the AC modes using the information $f_{i} \propto v_{i}n$, we first traced the refractive index changes with $V_{ext}$ by measuring the ellipsometric parameters, $\psi$ and $\Delta$. These parameters are given by the ratio of the Fresnel reflection coefficients $\rho_\pi$ and $\rho_\sigma$ for light polarized parallel and perpendicular to the incidence plane, respectively, as

\begin{equation}
\rho= \frac{\rho_{\pi}}{\rho_{\sigma}}=\tan(\psi)e^{i\Delta},
\end{equation}
where the wavelength ranged from 350 to 380 nm, and the incidence angle $\theta_0$ was 75$^{\circ}$.

With the measured parameters $\psi$ and $\Delta$ in Fig. 8(a--d), the complex refractive indices $\tilde{n}$ for both samples could be evaluated as
\small
\begin{equation}
\tilde{n}=\frac{\sqrt{1-4\sin^2(\theta_0)\tan(\psi)e^{i\Delta}+2\tan(\psi)e^{i\Delta}
+\tan^2(\psi)e^{i\Delta}}}{\sin(\theta_0)^{-1}\cos(\theta_0)[1+\tan^2(\psi)e^{i\Delta}]}, \end{equation}
\normalsize
where the real part of $\tilde{n}$ is the refractive index $n$. The unperturbed dielectric constant $\epsilon^{o}$ of sample A (sample S) was evaluated to be 8.64 (9.12) at 367.5 nm; using this value, the modulation of the permittivity $\Delta\epsilon (V_{ext})$ was subsequently extracted based on the relation $n=\sqrt{\epsilon^{o}+\Delta \epsilon (V_{ext})}$, as shown in Fig. 8(e) [Fig. 8(f)] for sample A (sample S).

%Figures 8(e) and 8(f) show the refractive indexes as a function of $V_{ext}$ at the probe wavelength of 367.5 nm for sample A and sample S, respectively.

\begin{figure}[!t]
\centering
\includegraphics[scale=0.2]{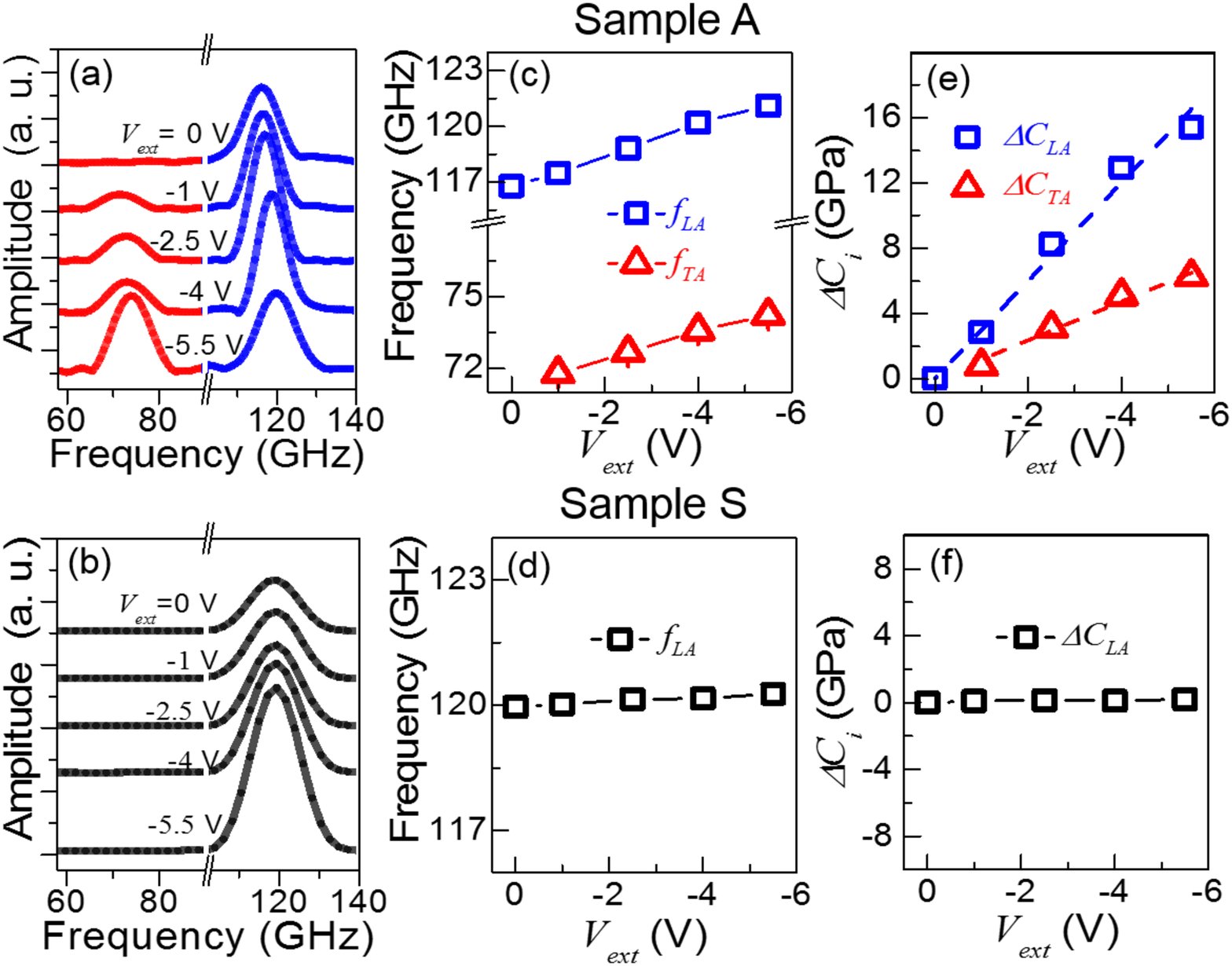}
\caption{FFT spectra for the ascending waves in (a) sample A and (b) sample S as a function of $V_{ext}$. Spectral peak frequencies of AC modes in (c) sample A and (d) sample S under different $V_{ext}$. $V_{ext}$-induced elastic constant variations in (e) sample A and (f) sample S.}
\end{figure}

On the other hand, the modal spectra of the DFP oscillations were traced without phase-change-induced errors by selecting temporal ranges for the FFT between $\tau_{i}/2$ and $\tau_{i}$, as shown in Fig. 9(a,b) for samples A and S, respectively. The peak frequencies of both the LA and TA modes in sample A increased by $\sim$1.7\% with increasing $V_{ext}$ in Fig. 9(c). In contrast, the spectral shift of the LA mode in sample S in Fig. 9(d) was relatively negligible, $\sim$0.25\%. Then, inserting the measured $n$ into $f_i$ under $V_{ext}$, we extracted $ v_{i}$ equal to  $\sqrt{ \frac{C_i+\Delta C_i (V_{ext})}{\rho_0}} $. The corresponding bias-dependent changes, $\Delta \epsilon (V_{ext})$ in Fig. 8(e,f) and $\Delta C_i (V_{ext})$ in Fig. 9(e,f), were further compared with the analytical expressions as $\Delta \epsilon \simeq p_{13} d_{33}E_z + \frac{1}{2}(p_{13}+ p_{15}) d_{15}E_x $, $\Delta C_{LA} \simeq C_{333}^H d_{33}E_z+\frac{(C_{355}^M)^2}{C_{33}^0-C_{44}^0}d_{15}^2E_x^2$, and $\Delta C_{TA}  \simeq C_{443}^H d_{33}E_z-\frac{(C_{355}^M)^2 }{C_{33}^0-C_{44}^0} d_{15}^2 E_x^2$. Here, we incorporated the optical birefringence and strain-induced nonlinear elasticity in the anisotropic region based on the photoelastic tensor component $p_{ij}$ and the TOE coefficient for the normal (shear) strain,  $C_{lmn}^H$ ($C_{lmn}^M$). %$C_{33}^0$ ($C_{44}^0$) is the unperturbed elastic constant for the LA (TA) mode.
The relatively moderate behavior of $f_{LA}$ with $V_{ext}$ in sample S implies that the shear strain contribution is dominant even in modifying $v_{LA}$, as demonstrated in Eq. (6).

\section{Role of symmetry in the detection of acoustic phonons }

In an optically isotropic medium such as $c$-GaN, dielectric tensor modulation due to shear strain can be detected only under oblique probe light incidence \cite{Nye,Matsudaoff}. The shear strain in a laterally biased anisotropic medium, on the other hand, induces perturbation not only in the off-diagonal components, but also in the diagonal component of the dielectric tensor. Accordingly, the TA mode was optically detectable even when the incidence angle was normal to the surface, as represented by the digitized appearance in Figs. 5(a) and 7. In this section, we experimentally verify the optical anisotropy and corresponding changes in the detection sensitivity by measuring the AC modal properties using different combinations of probe polarizations $\phi$ and incidence angles $\theta$.

\subsection{Electric-field-induced optical birefringence }

\begin{figure}[!t]
\centering
\includegraphics[scale=0.18]{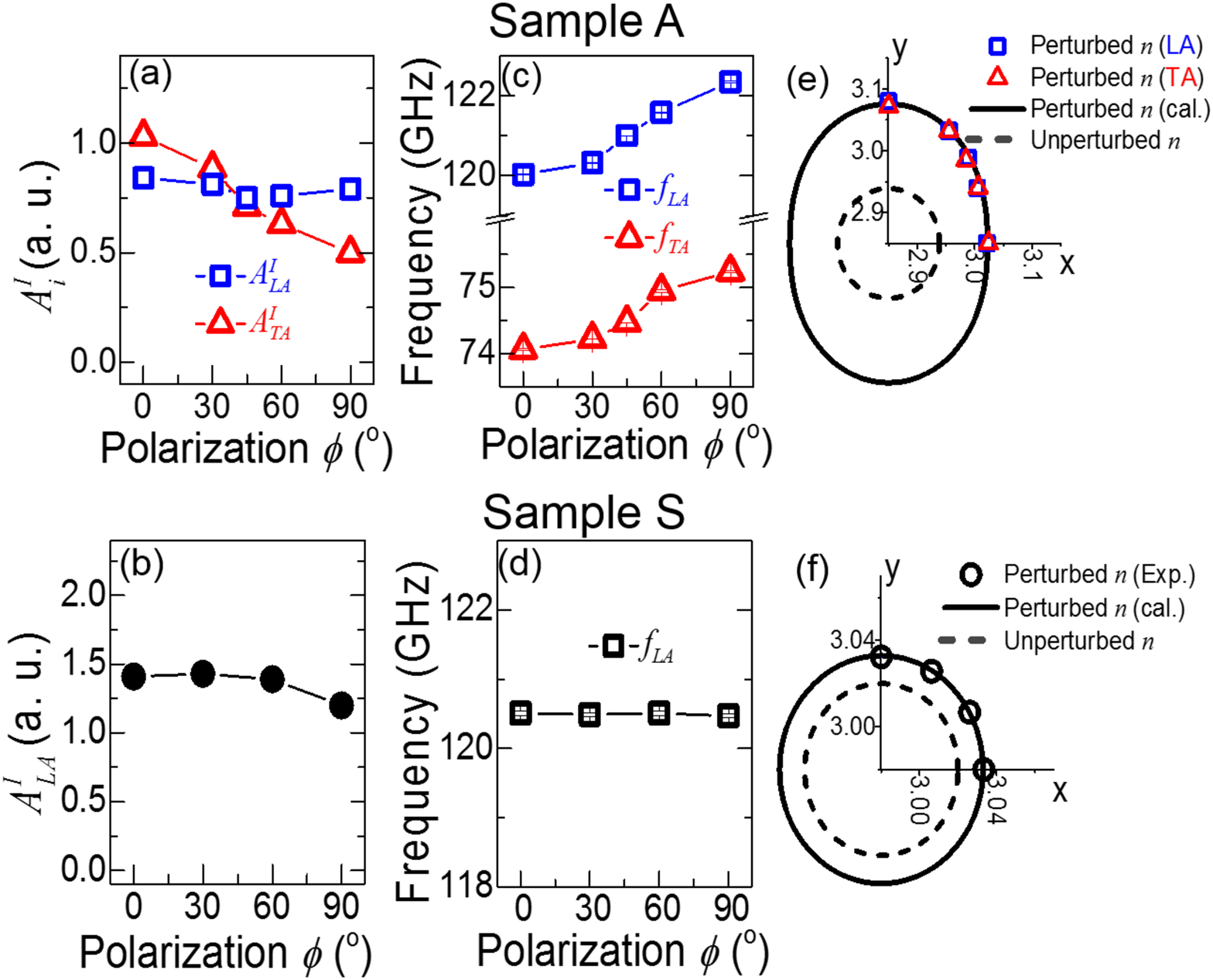} \caption{
AC mode amplitudes of (a) sample A and (b) sample S for $\phi$ ranging from $0^{\circ}$ to $90^{\circ}$ in time domain $I$. Peak frequency for each AC mode in (c) sample A and (d) sample S as a function of $\phi$. Refractive index ellipsoids for perturbed (at $-$5.5 V) and unperturbed (at 0 V) cases in (e) sample A and (f) sample S.}
\end{figure}

Regarding the electrically manipulated crystal symmetry, $E_x$ breaks the hexagonal symmetry of sample A via the TOE tensors into monoclinic symmetry \cite{Fuck}. Concretely, the tilting of the principal axis away from the $c$ axis by $E_x$-induced monoclinic deformation could be illustrated by angular perturbations of the elastic and dielectric tensors in the $x-y$ plane.

Figure 10(a,b) shows the spectral amplitudes of DFP oscillations from both samples at a $V_{ext}$ value of $-$5.5 V, integrated over $\tau_{i}/2<t<\tau_{i}$, as we rotate the probe polarization $\phi$ from the $x$ axis at $\theta=0$. In sample A, the TA amplitude ($A^I_{TA}$) monotonically decreased by about 50\% with increasing $\phi$, in clear contrast to the constant LA mode ($A^I_{LA}$) in Fig. 10(a), indicating that the TA mode was partially polarized along the same direction as $E_x$. Figure 10(b) shows that in sample S, $A^I_{LA}$ was invariant with respect to the probe polarization $\phi$, as can be inferred from Eq. (13).

Despite the large value of $V_{ext}$, the modal frequencies $f_{LA}$ and $f_{TA}$ in sample A exhibited very similar increases with $\phi$, as shown in Fig. 10(c), owing to the birefringence in the anisotropic region. In contrast, no notable change was observed for sample S, as shown in Fig. 10(d). As a function of $\phi$, in this regard, the $n$ value of sample A was extracted from either $f_{LA}$ (blue squares) or $f_{TA}$ (red triangles) in Fig. 10(e). The dashed inner circle in Fig. 10(e) displays the isotropic refractive index ellipsoid without $E_x$, whereas the solid outer circle exhibits the values calculated using the $E_x$-induced birefringence. In sample S, which has a symmetric potential distribution even under an applied $V_{ext}$, the $V_{ext}$-induced birefringence and nonlinear elasticity was not observed in Fig. 10(f). For  this reason, the hexagonal symmetry of the lattice structure should be preserved with $n_{\mathcal{E}^{\hbar \omega}_x}=n_{\mathcal{E}^{\hbar \omega}_y}$.  The comparison of the samples therefore confirms the dominant role of the lateral potential gradient in manipulating the crystalline symmetry, concurrently with the modified elastic and optical coefficients.

\subsection{Optical characterization of anisotropy in sample A}

\begin{figure}[!t]
\centering
\includegraphics[scale=0.15]{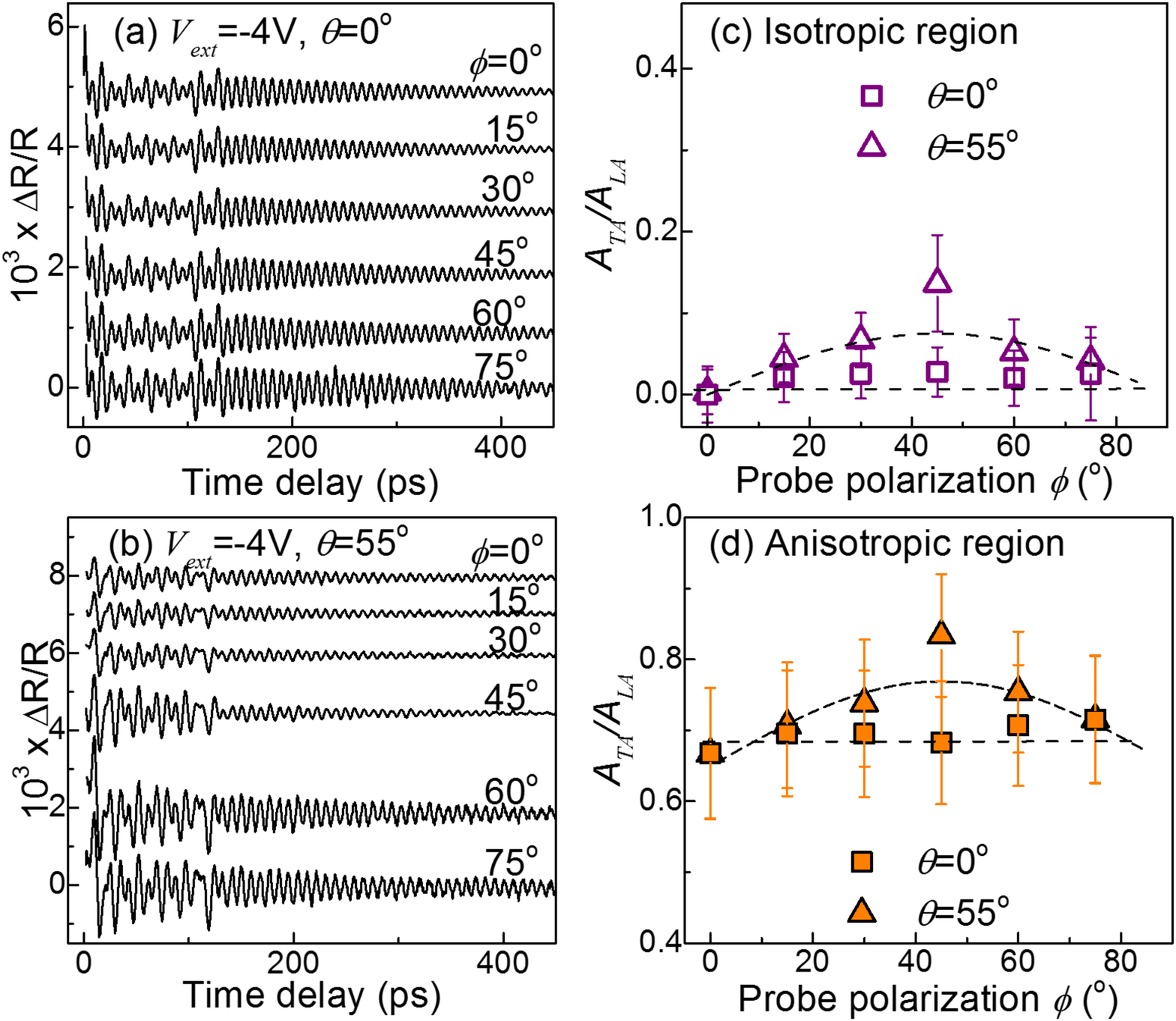}
\caption{DFP oscillations with variable probe polarization at (a) $\theta=0^{\circ}$ and (b) $\theta=55^{\circ}$ under $-$4 V. The $\phi$-dependent relative amplitudes of the TA mode, $A$$_{TA}$/$A$$_{LA}$, in the (c) isotropic and (d) anisotropic region at probe incidence angles of 0$^{\circ}$ and 55$^{\circ}$.}
\end{figure}

%experiment: incidence angle and polarization 45 degree and an analyzer for best detection of shear strain in an isotropic region.

To further investigate the optical anisotropy in sample A, we varied the probe polarization $\phi$ at a fixed $\theta$, now with an $s$-polarized analyzer in front of the probe detector. When the probe beam is incident on the sample at nonzero $\theta$, the DFP interference for the TA mode is expected to be maximized at a $\phi$ value of 45$^{\circ}$ in such a geometry\cite{Matsuda}. Since the PP signals became attenuated as they propagated downward toward the $n$-GaN side, the laser wavelength was further increased to 375 nm to ensure a deeper probe beam penetration depth $\xi$ ($\sim$1.5 $\mu$m). The $\phi$-dependent DFP oscillations under $-$4 V are shown in Fig. 11(a,b) for normal and 55$^{\circ}$ probe incidence, respectively. To extract the spectral amplitudes of AC modes propagating in regions of different symmetry, we performed FFTs in two distinct temporal ranges, that is, in the anisotropic region for $t<\tau_{LA}$ (spatial extent: 0 $<$ $z$ $<$ $l_D$ + 440 nm, where $l_D$ is the $n$-depletion region width) and in the isotropic region for $t>\tau_{TA}$ ($z$ $>$ $l_D$ + 440 nm). In Fig. 11(c,d), the relative amplitudes of the TA modes are traced with respect to $\phi$ in the isotropic and anisotropic regions, respectively.

%Normal  incidence
As can be expected intuitively in the isotropic region, TA mode propagation was not detected under normal incident probe light, as shown in Fig. 11(c); in contrast, in the anisotropic region, TA modes were observed consistently as $\phi$ changed, even for normal probe incidence in Fig. 11(d). The result reveals that the diagonal components of the dielectric tensor were perturbed not only by the LA modes but also by the TA modes, as indicated by Eq. (17). In other words, the biased region is no longer isotropic; consequently, both the LA and TA modes play roles in Maxwell's equation for normally incident light. The different contributions of the modes are described in terms of the strain distributions along the $z$ axis and the diagonal photoelastic coefficient for each mode.

%55 degree incidence
When the probe beam is obliquely incident on the sample, the DFP interference for a TA mode is thought to follow the change in $\Delta R/R \propto \sin(\phi)\cos(\phi)$, as denoted by the dashed line in Fig. 11(c) for an isotropic medium,  because $s$-polarized light is flipped into $p$-polarization upon reflection from the TA wavepackets and vice versa for nonzero $\theta$ \cite{Matsudaoff}.  The isotropy of the unbiased region was confirmed by the agreement with the conventional $\sin(\phi)\cos(\phi)$  dependence, except for a hump at 45$^{\circ}$. The deviation is possibly attributed to the reduced sensitivity of the LA mode due to the screening effect of the TA mode.  The result for the anisotropic region in Fig. 11(d), where the light reflection is now influenced by both the diagonal and off-diagonal perturbations of the TA modes, indicates that the off-diagonal effect of the TA waves on the probe polarization was boosted by the amount of anisotropy.

%Screening effect of TA mode on LA mode sensitivity}

\begin{figure}[!t]
\centering
\includegraphics[scale=0.195]{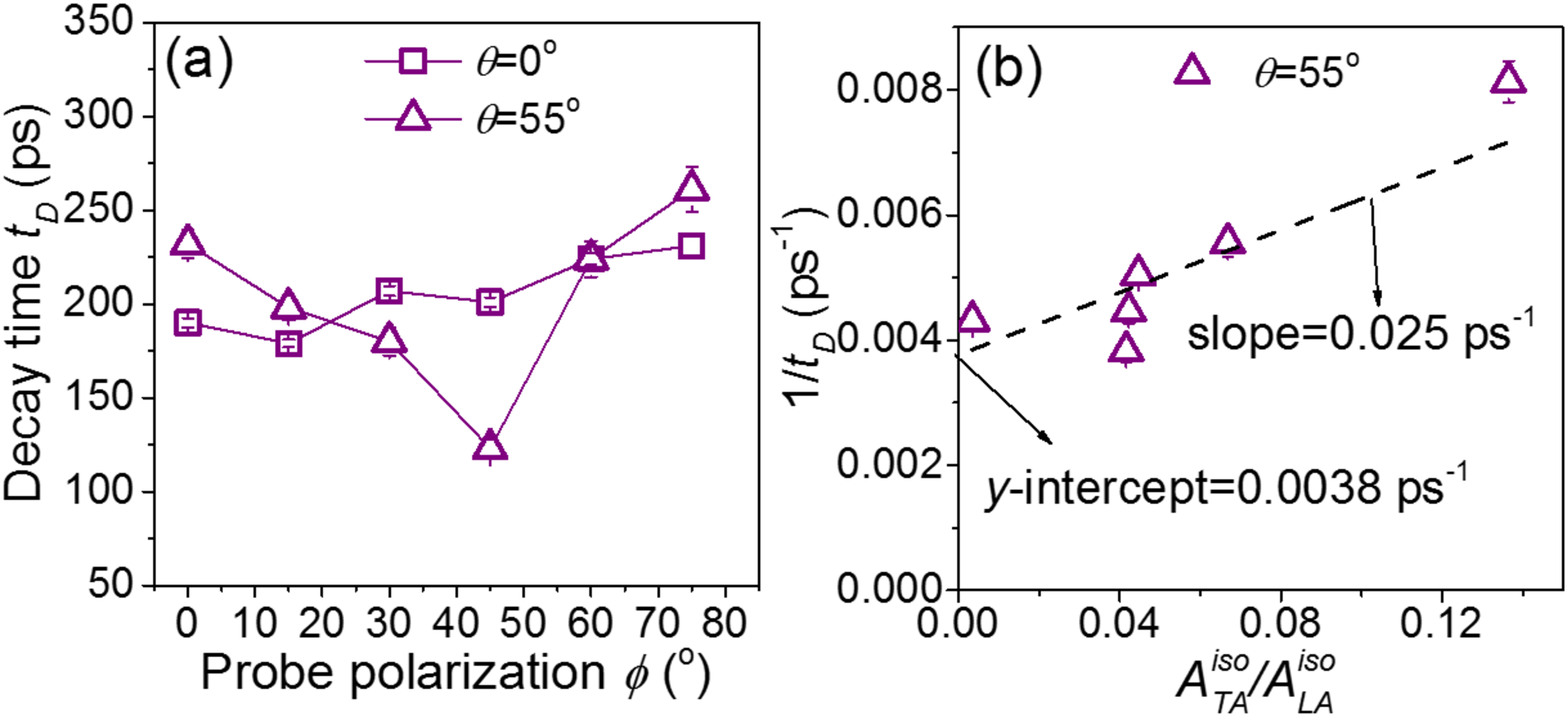}
\caption{(a) Decay times of LA mode signals in the isotropic region measured with normal (squares) and 55$^{\circ}$ (triangles) probe incidence. (b) Decay rate of LA mode signal at $\theta=55^{\circ}$ versus the relative amplitude of the TA mode.}
\end{figure}

The acoustic screening effect was substantiated by the close relationship between the decay time $t_D$ for the LA mode and the relative amplitude of the TA mode signal in the isotropic region, as shown in Fig. 12(a). For normal probe incidence, where the TA mode was not detected, $t_D$ was held at $\sim200$ ps. On the other hand, the behavior of $t_D$ was opposite to that of $\frac{A_{TA}}{A_{LA}}$ with respect to $\phi$ for $\theta=55^{\circ}$. To further specify the relationship, the decay rate $1/t_D$ for a 55$^{\circ}$ probe incidence is plotted as a function of  $\frac{A_{TA}}{A_{LA}}$  in Fig. 12(b). The $y$ intercept, 0.0038 ps$^{-1}$, coincided with the decay rate resulting from the optical absorption with a skin depth of $\sim$2 $\mu$m. The slope, 0.025 ps$^{-1}$, revealed the screening effect of the TA mode on LA mode detection as $\Delta R/R = \int dz [(\Delta \Gamma_{TA}) ^2 \digamma_{LA}] e^{-z/\xi}  \eta_{LA} \propto \int dz [e^{-\gamma z} \digamma_{LA}] e^{-z/\xi} \eta_{LA}$, where the empirical decay factor $\gamma$ is $\frac{0.025A_{TA}}{v_{LA}A_{TA}}$.

\subsection{External manipulation of the detection sensitivity in sample A}

\subsubsection{Experimentally measured AC sensitivities in regions of different symmetry}

To gain a more quantitative insight into the relationship between $\digamma_i$ and the axial symmetry, we selectively measured the $s$-polarized component of the probe reflectance as a function of $\theta$ at $-$4 V, as shown in Fig. 13. The probe polarization angle $\phi$ was fixed at 45$^{\circ}$ to maximize the sensitivity of the $x$-polarized TA mode.  As a result, more complicated AC propagation dynamics appear in Fig. 13; the TA mode signals monotonically increased with $\theta$ not only in time domain $I$ but also in time domain $III$. The abruptly suppressed TA mode at $\tau_{TA}$ was particularly prominent at $\theta$=$0^{\circ}$. In contrast, the LA amplitude with respect to $\theta$ varied in different time domains, slightly increasing up to $\tau_{LA}$ and then rapidly decreasing after 2$\tau_{LA}$. In an intermediate time domain, $II$, the increasing and decreasing tendencies were mixed for LA waves.

\begin{figure}[!t]
\centering
\includegraphics[scale=0.35]{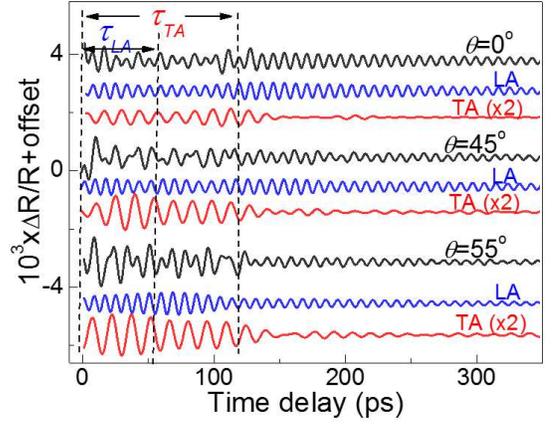}
\caption{ Time domain signals (black line) from sample A for different probe incidence angles were filtered into the LA (blue line) and TA mode (red line) frequencies.}
\end{figure}

The Fourier-transformed amplitudes $A^j_i$ for mode index $i$ were taken in the time domains $j$=\{$I,II,III$\}.  We assumed that the spatially concentrated AC strains are simplified to $\eta_{i}^{a}$=$\int_{D_i^{a}}\varepsilon_{i}(z)dz\cdot\delta(z+v_{i}t)$ for the ascending waves and $\eta_{i}^{d}$=$\int_{D_i^{d}}\varepsilon_{i}(z)dz\cdot\delta(z-v_{i}t)$ for the descending waves, where $\varepsilon_{LA}(z)$ ($\varepsilon_{TA}(z)$) corresponds to the normal (shear) strain, and the region of integration, $D_i^{d}$  ($D_i^{a}$), corresponds to the SDR (the $i$ region for the LA mode or the $n$-depletion region for the TA mode).

The ascending AC components, $\eta_{i}^{a}$, propagate toward the surface  in time domain $I$, as described in Fig. 14(a); they then propagate back to the substrate after being reflected from the $p$-GaN/ITO interface during time domain $II$ in the anisotropic region, and finally enter the isotropic region in time domain $III$. On the other hand, the descending components, $\eta^{d}_{i}$, which were generated in the SDR, traverse the anisotropic region in time domain $I$ and propagate in the isotropic region toward the substrate when $t>\tau_{i}$. Therefore, $A^{j}_{i}$, which was experimentally extracted as shown in Fig. 14(b--g), could be further analyzed as
\begin{equation} \begin{array}{l l}
A^{I}_{i}=\kappa^{I}_i\digamma_{i}^{ani}\eta_{i}^{a}+\kappa^{I}_i \digamma_{i}^{ani}\eta_{i}^{d}
& \quad {(0<t<\tau_{i}),}\\ \\
A^{II}_{i}=\kappa^{I}_ir \digamma_{i}^{ani} \eta_{i}^{a}+\kappa^{II}_i \digamma_{i}^{iso} \eta_{i}^{d}
& \quad {(\tau_{i}<t<2\tau_{i}),}\\ \\
A^{III}_{i}=\kappa^{II}_i r \digamma_{i}^{iso} \eta_{i}^{a}+\kappa^{III}_i \digamma_{i}^{iso} \eta_{i}^{d}
& \quad {(2\tau_{i}<t<3\tau_{i}),} \end{array} \end{equation}
where $\digamma_{i}^{ani}$ ($\digamma_{i}^{iso}$) is the optical sensitivity of the AC modes for the anisotropic (isotropic) region, and $r$ is the AC mode reflectivity at the $p$-GaN/ITO interface.
$\kappa^{j}_i$=$\frac{\xi}{v_{i}\tau_{i}}$(1-$e^{-\frac{v_{i}\tau_{i}}{\xi}}$)$e^{-(j-1)\frac{v_{i}\tau_{i}}{\xi}}$ is the probe light attenuation factor based on the wavelength-dependent $\xi$, where $j$=$\{I,II,III\}\equiv \{1,2,3 \}$. The sensitivity $\digamma_{i}^{ani}$ ($\digamma_{i}^{iso}$) was expressed in terms of $A_{i}^{I}$ ($A_{i}^{III}$) and then substituted into the right-hand side of the second equation in Eq. (25) to determine the optimal values of $x=\eta_{i}^{a}/(\eta_{i}^{a}+\eta_{i}^{d})$, which minimize the least-squares error $\delta(x)$ as
\begin{equation}
\delta(x)=\vert A_{i}^{II}-\kappa^{I}_ir\alpha_{i}\cdot
x\eta_{i}-\kappa^{II}_i\beta_{i}\cdot (1-x)\eta_{i}\vert^{2} \ \
\ \ \ (0<x<1), \end{equation}
where $\eta_{i}=\eta_{i}^{a}+\eta_{i}^{d}$. On the basis of the least-squares fitting by Eq. (26), $x$ was evaluated to be $\sim0.7$ ($\sim0.6$) for the LA (TA) mode, from which $\digamma_{i}^{ani}$ and $\digamma_{i}^{iso}$ could be experimentally extracted, as shown in Fig. 15(a--d).

\begin{figure}[!t]
\centering
\includegraphics[scale=0.46]{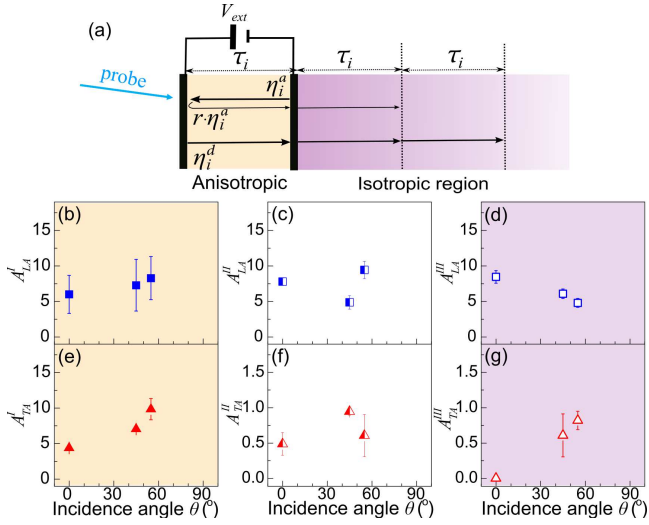}
\caption{(a) Propagation configurations of ascending ($\eta^{a}_{i}$) and descending ($\eta^{d}_{i}$) AC components in regions of different symmetry. (b--g) Fourier-transformed amplitudes of the DFP oscillations for the LA (b--d) and TA (e--g) modes in the time domain $j$=$\{I,II,III\}$.}
\end{figure}

\subsubsection{Electrically manipulated AC modal sensitivity }

The analytic expressions for $\digamma_{i}$ have been discussed separately for isotropic \cite{Matsudaoff} and anisotropic crystals \cite{Pezeril} in terms of the perturbations of the dielectric tensor due to strain. In isotropic materials, only the diagonal (off-diagonal) components in the perturbed dielectric tensor are induced by the LA (TA) waves. Subsequently,
%the previously investigated AC sensitivity expressions in the isotropic region could be adapted for our structure as $\digamma_{LA}^{iso} \propto \digamma^{dia}$ and $\digamma_{TA}^{iso} \propto \digamma^{off}$. In this simplified expression,
the opposite tendencies are predicted for $\digamma_{LA}^{iso}$ (dashed line) and $\digamma_{TA}^{iso}$ (dotted line) within our range of $\theta$ values, in agreement with the experimental values (empty squares and triangles) in Fig. 15(a,b).
\begin{figure}[!t]
\centering
\includegraphics[scale=0.12]{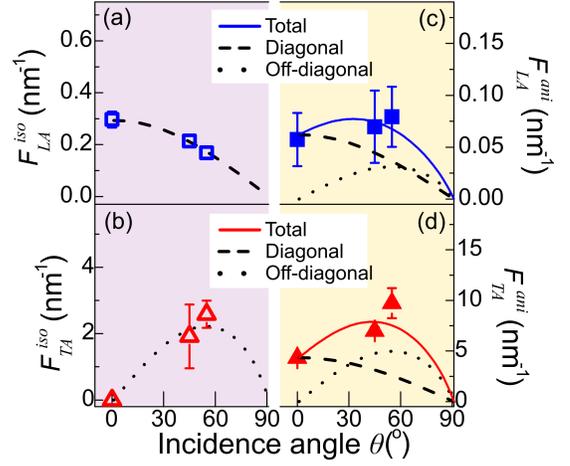}
\caption{Optical sensitivity of AC modes as a function of incident angle in the (a,b) isotropic and (c,d) anisotropic regions of sample A.}
\end{figure}

On the other hand, to calculate $\digamma_{i}^{ani}$, it is necessary to consider the mixed nature of the dielectric tensor modulations resulting from both diagonal and off-diagonal perturbations with empirically measured weighting factors to fit the experimental results in Fig. 15 (c,d) as
\begin{equation}
\begin{pmatrix}
\digamma_{LA}^{ani}\\
\digamma_{TA}^{ani}\\
\end{pmatrix}
=
\begin{pmatrix}
0.22&0.014\\
14.6&2.25\\
\end{pmatrix}
\begin{pmatrix}
\digamma_{LA}^{iso}\\
\digamma_{TA}^{iso}\\
\end{pmatrix}.
\end{equation}
Indeed, the experimentally obtained values of $\digamma_{i}^{ani}$ in Fig. 15(c,d) were reproduced by linear combinations of the diagonal perturbations in the dielectric tensor ($\propto \digamma_{LA}^{iso}$) and off-diagonal perturbations in the dielectric tensor ($\propto \digamma_{TA}^{iso}$). Most importantly, the digitized appearance of the typically forbidden TA mode at $\theta=0^{\circ}$ (Fig. 5 and 7) was theoretically verified by comparing the zero $\digamma_{TA}^{iso}$ [Fig. 15(b)] and nonzero value of $\digamma_{TA}^{ani}$ [Fig. 15(d)].

Although further detailed investigations of the vectorial influence of external fields and the corresponding carrier--phonon interactions are necessary, it is evident that the mixed contributions of both diagonal and off-diagonal modulations in the laterally biased region resulted in a gradual increment of $\digamma^{ani}_{LA}$ with increasing $\theta$ and allowed the TA mode detection, even at $\theta=0^o$ (which was otherwise forbidden\cite{Matsuda, Nye}), demonstrating the electrical controllability of axial symmetry in piezoelectric diode structures.

\section{Summary}

In summary, we investigated the influence of the potential distribution on AC modal behaviors in isotropic structures. The crystal symmetry was broken by lateral electric fields in the structure with asymmetric potential gradients, which enabled spatial manipulation of the selection rules and amplitudes of propagating acoustic waves. A detailed analysis of the propagation dynamics of the TA mode revealed abrupt suppression underneath the laterally biased depletion region, in striking contrast to the electrically independent appearance of the LA mode.  The results demonstrate the possibility of direct and active control of the acoustic modes in semiconductors simply by electrically manipulating the crystal symmetry. These findings could provide a novel degree of freedom in engineering the lattice vibrations in crystalline heterostructures.

\appendix*
\begin{appendices}
\section{Elastic and photoelastic tensors under different symmetries}
\label{sec:app}
As introduced in Sec. IV, the elastic tensor $C_{lm}$ of $c$-GaN was perturbed by externally applied strains, $\varepsilon_{3}$ ($\varepsilon_{zz}$) and $\varepsilon_{5}$ ($\varepsilon_{zx}$), as $C_{lm}=C_{lm}^0+\Delta C_{lm}=C_{lm}^0+C_{lmn}\varepsilon_{n}$, where $C_{lm}^0$ is the unperturbed elastic constant, and $C_{lmn}$ is the TOE tensor \cite{Fuck}. Analogous results for the photoelasticity could be deduced based on the geometric symmetry of the strained crystal \cite{Nye,Fuck} as $p_{lm}=p_{lm}^0+\Delta p_{lm}=p_{lm}^0+p_{lmn}\varepsilon_{n}$, where $p_{lm}^0$ is the unperturbed photoelastic constant, and $p_{lmn}$ is the the tensor components for the nonlinear photoelasticity. In Table I, the elastic and photoelastic tensors under different symmetry conditions are comparatively summarized.
% in agreement with the invariance of their components under certain rotations of the coordinate system

\begin{widetext}
\begin{table*}[!ht]
%\resizebox{\textwidth}
 \label{tensors}
\caption{Matrix representations of the elastic and photoelastic tensors with different symmetries.}
\setlength{\tabcolsep}{8pt}
\renewcommand{\arraystretch}{1}

\begin{tabular}{m{3cm} c  c}
\hline\hline %inserts double horizontal lines
\centering
Symmetry & Elastic tensor  & Photoelastic tensor  \\  % inserts table heading,
%\\ \includegraphics [scale=0.3]{FigureA2}  \\ \includegraphics [scale=0.3]{FigureA3}
\hline  \\  % inserts single horizontal line
\centering   Hexagonal \\ \includegraphics [scale=0.3]{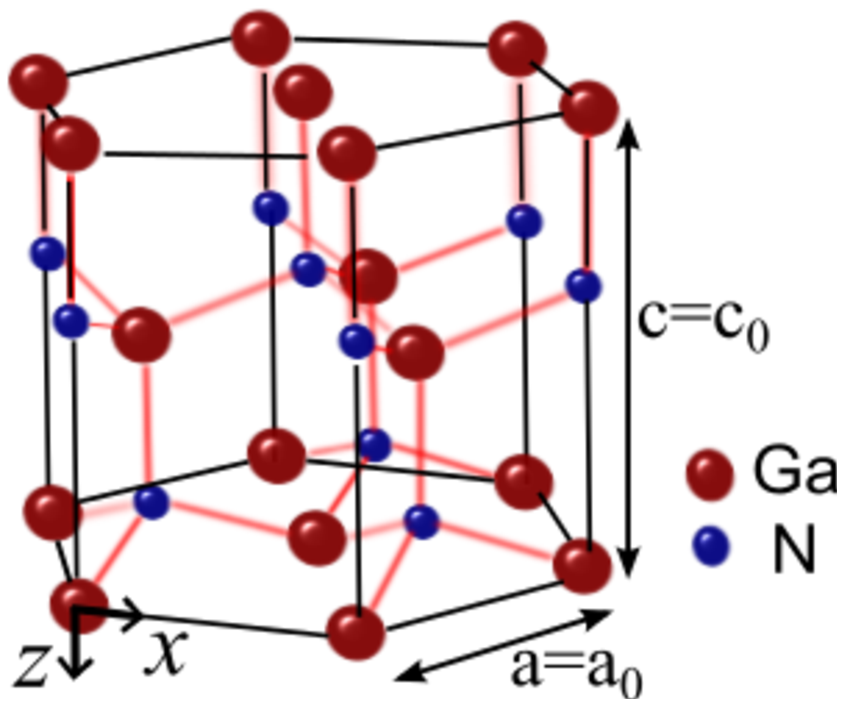} &  $[C^0]=\begin{bmatrix} C_{11}^0&C_{12}^0&C_{13}^0&0&0&0\\ C_{12}^0&C_{11}^0&C_{13}^0&0&0&0\\ C_{13}^0&C_{13}^0&C_{33}^0&0&0&0\\ 0&0&0&C_{44}^0&0&0\\ 0&0&0&0&C_{44}^0&0\\ 0&0&0&0&0&C_{66}^0\\ \end{bmatrix}$ & $[p^0]=\begin{bmatrix} p_{11}^0&p_{12}^0&p_{13}^0&0&0&0\\ p_{12}^0 & p_{11}^0&p_{13}^0&0&0&0\\ p_{13}^0&p_{13}^0&p_{33}^0&0&0&0\\ 0&0&0&p_{44}^0&0&0\\ 0&0&0&0&p_{44}^0&0\\ 0&0&0&0&0&p_{66}^0\\ \end{bmatrix}$ \\ [1.5cm] % inserting body of the table
\hline  \\
\centering  Hexagonal with $\varepsilon_{zz}$ \\ \includegraphics [scale=0.3]{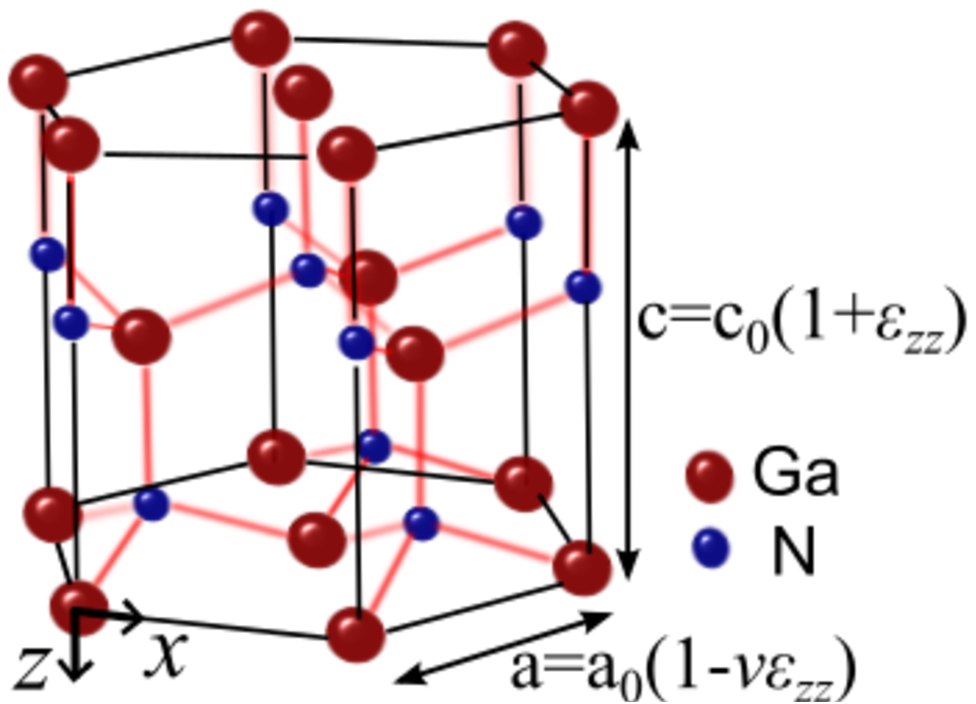} & $[C^0]+\begin{bmatrix} C_{113}^H& C_{123}^H& C_{133}^H&0&0&0\\ C_{123}^H& C_{113}^H& C_{133}^H&0&0&0\\ C_{133}^H& C_{133}^H& C_{333}^H&0&0&0\\ 0&0&0& C_{443}^H&0&0\\ 0&0&0&0& C_{443}^H&0\\ 0&0&0&0&0& C_{663}^H\\ \end{bmatrix} \varepsilon_{zz}$ & $[p^0]+\begin{bmatrix} p_{113}^H& p_{123}^H& p_{133}^H&0&0&0\\ p_{123}^H& p_{113}^H& p_{133}^H&0&0&0\\ p_{133}^H& p_{133}^H& p_{333}^H&0&0&0\\ 0&0&0& p_{443}^H&0&0\\ 0&0&0&0& p_{443}^H&0\\ 0&0&0&0&0& p_{663}^H\\ \end{bmatrix} \varepsilon_{zz}$\\[1.5cm]
\hline  \\

\centering Monoclinic with $\varepsilon_{zx}$ \\ \includegraphics [scale=0.3]{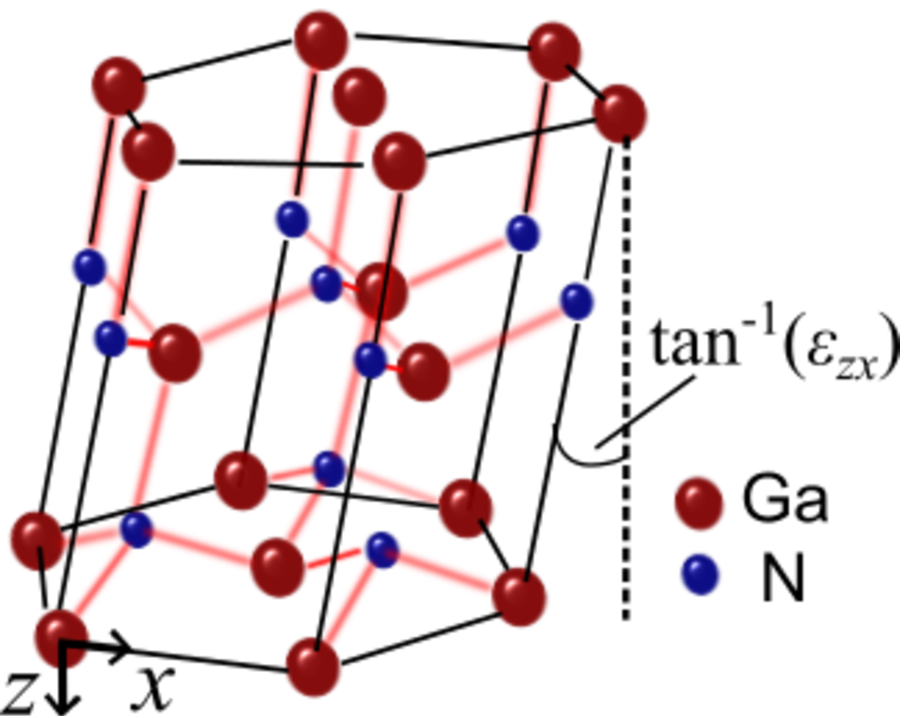}& $[C^0]+\begin{bmatrix}  0& 0& 0&0&C_{155}^M &0 \\  0& 0& 0&0&C_{255}^M &0 \\  0& 0& 0&0&C_{355}^M &0 \\ 0&0&0&0&0&C_{465}^M\\ C_{155}^M&C_{255}^M&C_{355}^M&0&0&0\\ 0&0&0&C_{465}^M&0&0\\ \end{bmatrix}\varepsilon_{zx}$ &  $[p^0]+\begin{bmatrix}  0& 0& 0&0&p_{155}^M &0 \\  0& 0& 0&0&p_{255}^M &0 \\  0& 0& 0&0&p_{355}^M &0 \\ 0&0&0&0&0&p_{465}^M\\ p_{155}^M&p_{255}^M&p_{355}^M&0&0&0\\ 0&0&0&p_{465}^M&0&0\\ \end{bmatrix}\varepsilon_{zx}$  \\[1.5cm]
\hline \hline %inserts single line \includegraphics[width=0.3\textwidth, height=60mm]{Figure6}
\end{tabular}
%\label{table:symmetry} % is used to refer this table in the text
\end{table*}
\end{widetext}

\section{derivation of formulas}
The AC velocities for LA- and TA-mode respectively are obtained by solving Eq.(3) and simplified under the assumption that (i) $C_{lm}^0 \gg \Delta C_{lm}$ and (ii) $C_{333}^M \sim C_{443}^M$,
%\begin{widetext}
\small
\begin{eqnarray}
  \label{eq:A-1}
 v_{LA} &=&\frac{\sqrt{C_{33}+C_{44}+\sqrt{(C_{33}-C_{44})^2 +4C_{35}^2}}}{\sqrt{2\rho_0}} \\\nonumber
&\simeq& \frac{\sqrt{C_{33}+\frac{C_{35}^2}{C_{33}-C_{44}}}}{\sqrt{\rho_0}}\\\nonumber
&\simeq& \frac{\sqrt{C_{33}^0+C_{333}^H d_{33}E_z+\frac{(C_{355}^M)^2 }{C_{33}^0-C_{44}^0} d_{15}^2 E_x^2}}{\sqrt{\rho_0}}
\end{eqnarray}
and
\begin{eqnarray}
\label{eq:A-2}
v_{TA\parallel x}&=&\frac{\sqrt{C_{33}+C_{44}-\sqrt{(C_{33}-C_{44})^2 +4C_{35}^2}}}{\sqrt{2\rho_0}}\\ \nonumber
&\simeq& \frac{\sqrt{C_{44}-\frac{C_{35}^2}{C_{33}-C_{44}}}}{\sqrt{\rho_0}}\\\nonumber
&\simeq& \frac{\sqrt{C_{44}^0+C_{443}^H d_{33}E_z-\frac{(C_{355}^M)^2 }{C_{33}^0-C_{44}^0} d_{15}^2 E_x^2}}{\sqrt{\rho_0}},
\end{eqnarray}
\normalsize
%\end{widetext}
where Eq.~(\ref{vla}) and Eq.~(\ref{vta}) in Sec.IV are the final form.

%The photoelastic tensor, $p_{ij}$, in Eq.(10) for the given symmetry has the following form:
%\begin{multline}
%\label{p_ij}
%  p_{ij} =
%\begin{bmatrix}
%p_{11}&p_{12}&p_{13}&0&p_{15}&0\\
%p_{12}&p_{11}&p_{13}&0&p_{25}&0\\
%p_{13}&p_{13}&p_{33}&0&p_{35}&0\\
%0&0&0&p_{44}&0&p_{46}\\
%p_{15}&p_{25}&p_{35}&0&p_{44}&0\\
%0&0&0&p_{46}&0&p_{66}\\
%\end{bmatrix}
%\end{multline}

%In Eq.(21) of Sec.V,
%the elastic constants, $C_{ij}$ has the form in the Wurzitze structure,
%\begin{multline}
%\label{eq:C-WZ}
%  C_{ij} =
%\begin{bmatrix}
%C_{11}&C_{12}&C_{13}&0&0&0\\
%C_{12}&C_{11}&C_{13}&0&0&0\\
%C_{13}&C_{13}&C_{33}&0&0&0\\
%0&0&0&C_{44}&0&0\\
%0&0&0&0&C_{44}&0\\
%0&0&0&0&0&C_{66}\\
%\end{bmatrix}
%\end{multline}

%\begin{multline}
% \label{eq:dmatrix}
%d_{ij} =
%\begin{bmatrix}
%0&0&d_{31}\\
%0&0&d_{32}\\
%0&0&d_{33}\\
%0&d_{24}&0\\
%d_{15}&0&0\\
%0&0&0\\
%\end{bmatrix}
%\end{multline}
\end{appendices}
\begin{acknowledgments}
We acknowledge useful discussions with Y. S. Lim and K. S. Kyhm. This work was funded by Samsung Electronics (SRFC-IT1402-07).
\end{acknowledgments}

\end{document}